\documentclass[12pt]{article}
\usepackage{amsfonts}
\usepackage{amsmath}
\usepackage{amssymb}
\usepackage{color}
\usepackage{graphicx}
\usepackage[font=small,format=plain,labelfont=bf,up,textfont=it,up]{caption}

\makeatletter
\@addtoreset{equation}{section} 
\makeatother

\def\appendix#1
{
\addtocounter{section}{1}\setcounter{equation}{0}
 \renewcommand{\thesection}{\Alph{section}}
 \section*{
 \thesection\protect\indent \parbox[t]{16.715cm}{#1}}
 \addcontentsline{toc}{section}{Appendix \thesection\ \ \ #1}
}
\begin{document}
\begin{center}
\textbf{\Large Emergence of Homogeneous and Isotropic Loop Quantum Cosmology from Loop Quantum Gravity: Lowest Order in $h$}
\end{center}
\begin{center}\textbf{Chun-Yen Lin}\\ \textit{Physics Department, University of California\\ Davis, CA 95616}\\ 
\end{center}

{\small 
\begin{center}\textbf{Abstract}
\end{center}
To derive loop quantum cosmology from loop quantum gravity, I apply the model given in \cite{lin1} to a system with coupled gravitational and matter fields. The matter sector consists of a scalar field $\phi$ serving as a cosmological clock, and other fields $\{\psi\}$ providing physical spatial coordinates and frames. The physical Hilbert space of the model is constructed from the kinematical Hilbert space of loop quantum gravity, and the local observables in the physical Hilbert space are constructed using the matter coordinates and frames. A specific coherent physical state is then chosen, whose expectation values of the local observables give rise to homogeneous, isotropic and spatially flat gravitational and $\phi$ fields at a late clock time. The equations governing these fields may be derived using the symmetry of the physical Hilbert space.  When the matter back reactions from $\{\psi\}$ are negligible, the result gives a specific loop quantum cosmological model in the $O(\hbar^0)$ approximation, with calculable higher order corrections.

\section{Introduction}

 Loop quantum cosmology refers to a set of quantum cosmological models \cite{lqc0,lqcint1,lqcint2,lqcint3,lqcint4} incorporating key features of loop quantum gravity \cite{intro1,intro,perez}. As a candidate fundamental theory of gravity, loop quantum gravity has a rigorously defined kinematical Hilbert space describing quantum spatial geometry that is discretized in Planck scale. Building on this kinematical foundation, theorists are now tackling the challenge of obtaining the dynamics. By introducing the key features of loop quantum gravity into symmetric cosmological models, loop quantum cosmology aims to explore the effects of the quantum geometry in cosmic evolution.

 Loop quantum cosmology bypasses much of the complexity in loop quantum gravity because of symmetry reductions, and its dynamics is well-defined and extensively studied. However, the simplifications come with ambiguities: in the same physical setting, different symmetry reduction procedures can lead to distinct loop quantum cosmological models. Thus, several models in loop quantum cosmology have been proposed based on different physical considerations. However, for an isotropic, homogeneous and $k\geq 0$ universe with massless scalar matter fields, several robust features prevail in the family of models \cite{lqc0,lqcint1,lqcint2,lqcint3,lqcint4}: 1) they reproduce classical FRW cosmology in their large-scale semi-classical limits; 2) the semi-classical limits deviate from FRW cosmology near the initial singularity, replacing it with a well-behaved bounce of the scale factor \cite{lqcimp3,lqcimp4}; 3) with a proper coherent state, an in-built slow-roll inflation phase occurs following the bounce \cite{lqcimp1,lqcimp2}. These results are obtained through both numerical and analytical approaches to different degrees of approximation. Moreover, the study of  the cases with either negative $k$, inhomogeneity or anisotropy \cite{lqc0,lqc1,aniso1,aniso2,aniso3} also finds a bounce in Bianchi I, II and IX anisotropic models.

 With its rich implications, it is essential for loop quantum cosmology to find its root in loop quantum gravity, so that the implications can be attributed to fundamental principles. On the other hand, people are searching for observable predictions in loop quantum gravity, hoping to derive loop quantum cosmology.
 
Aiming at these goals, I apply the model proposed in \cite{lin1} to a cosmological system with $\Lambda=0$. The matter sector of the system consists of a chargeless and massless scalar field $\phi$ and other matter fields denoted as $\{\psi\}$. These matter fields live in the quantum geometry of space given by the gravitational sector of the system. The model describes this system using the kinematical Hilbert space of loop quantum gravity, called knot space.

Based on knot space, the physical Hilbert space $\mathbb H$ of the model is built under concrete assumptions. Local observables in $\mathbb H$ for gravitational and $\phi$ fields are constructed, using spacetime coordinates provided by the matter fields \cite{kuchar,torre}. Specifically, the $\phi$ field serves as the cosmological clock, and the fields in $\{\psi\}$ provide the spatial coordinates and frames. For a suitable coherent state in $\mathbb H$, the expectation values of these observables give rise to emergent classical gravitational and $\phi$ fields in the spacetime manifold. In our context, we will choose a coherent state that gives homogeneous, isotropic and spatially flat emergent fields at an early clock time. The cosmic evolution of the emergent fields will  be calculated using the symmetry of $\mathbb H$. When the matter back reactions from $\{\psi\}$ sector are negligible, the result conforms with a specific loop quantum cosmological model at order $O(\hbar^0)$. To the approximation, it recovers FRW cosmology in large scales and resolves initial singularity by the big bounce.

\section{Kinematic Hilbert Space of Loop Quantum Gravity }

This section is a brief introduction to the kinematics of loop quantum gravity including  gravitational and matter sectors. The kinematical Hilbert space, called knot space, gives a background independent description of the spatial quantum geometry and the resident matter fields.

\subsection{Ashtekar Variables for Gravitational and Matter Fields }
Loop quantum gravity is based on a canonical general relativity in the Ashtekar formalism \cite{form1,form2,form3}. This formalism describes gravitational fields in a form similar to that of matter gauge fields.

Traditionally, canonical general relativity uses spatial metric and extrinsic curvature defined in the spatial manifold $M$ as phase space variables \cite{form1,form2,form3}. As an alternative, triad fields consisting of three orthonormal vector fields $\{e^a_i (\text{x})\} (\text{x}\equiv (x,y,z); a=x,y,z;  i=1,2,3)$ can be used in place of the spatial metric. 
In the basis of $e^a_i (\text{x})$ and its inverse $e_a^i (\text{x})$, the spatial Levi-Civita connection and extrinsic curvature take the forms $\Gamma ^i_a (\text{x})$ and $K^i_a(\text{x})$. Ashtekar variables $(  A^i_a(\text{x}),E^a_i (\text{x}))$ are related to $(e^a_i (\text{x}), K^i_a(\text{x}))$ through a canonical transformation by \cite{form2}\cite{form3}:
\begin{equation}
\begin{split}
E^a_i (\text{x})\equiv {\det (e)} e^a_i(\text{x});\rule{10pt}{0pt}
A^i_a(\text{x}) \equiv \Gamma ^i_a (\text{x})+ \gamma K^i_a(\text{x})
\end{split}
\end{equation}
where the real number $\gamma$ is called the Immirzi parameter. By construction, the fields $E^a_i(\text{x})$ are densitized triad fields, and the fields $A^i_a(\text{x})$ are $SO(3)$ gauge fields. The variables have the non-vanishing Poisson brackets:
\begin{equation}
\begin{split}
\{A^i_a(\text{x}), E^b_j(\text{y})\}= 8\pi( G/c^3 )\gamma \delta^b_a \delta^i_j \delta (\text{x},\text{y})
\end{split}
\end{equation}
where $G$ is Newton's constant. Note that one may replace the $SO(3)$ symmetry group with $SU(2)$ in this formalism, since $SU(2)$ and $SO(3)$ share the same Lie algebra.

Matter fields with a gauge group $\mathcal G$ can be included in this formalism \cite{matter1,matter2,thiemann}. Using the triad basis, we describe the fermion, scalar and gauge fields by $(\xi^{\bar i}_{\bar{\text i}}(\text{x}),\pi_{\bar i}^{\bar{\text i}}(\text{x}))$, $(\phi^{\text i}(\text{x}),P_{\text i}(\text{x}))$ and $\underbar{A}^{\text i}_a(\text{x}),\underbar E_{\text i}^a(\text{x}))$. Here $\bar i$, ${\bar{\text i}}$ and ${\text i}$ are respectively (gravitational spin $ \frac{1}{2}$) $SU(2)$, $\mathcal G$ and adjoint $\mathcal G$ indices.

In this formalism, the action of general relativity possesses the following local symmetries: 1) time re-foliation invariance, which introduces the Hamiltonian constraint $H(\bar N)=0$; 2) spatial diffeomorphism invariance, which introduces the momentum constraint $M(\bar V)=0$; 3) local $SU(2)$ invariance, which introduces the $SU(2)$ Gauss constraint $G(\bar{\Lambda})=0$; 4) local $\mathcal G$, invariance which introduces the $\mathcal G$ Gauss constraint $\underbar{G}(\bar{\lambda})=0$. Each constraint is a functional of the dynamical fields and the Lagrangian multipliers $\bar N(\text{x})$, $\bar V^a(\text{x})$, $\bar{\Lambda}^i(\text{x})$, $\bar{\lambda}^{\text i}(\text{x})$ (the bars indicate their non-dynamical nature). The first three constraints result from spacetime diffeomorphism symmetry, and they consist of pure gravitational and matter terms:
\begin{equation}
\begin{split}
H (\bar N)= H_g (\bar N)+ H_m (\bar N);\rule{10pt}{0pt}
G(\bar{\Lambda})= G_g(\bar{\Lambda})+ G_m(\bar{\Lambda});\rule{10pt}{0pt}
M(\bar V)= M_g(\bar V)+ M_m(\bar V)
\end{split}
\end{equation}
 Here, we give the explicit forms of the pure gravitational terms, denoting $\kappa\equiv 8\pi( G/c^3 )$:
\begin{equation}
\begin{split}
H_g (\bar N)\equiv\int_M d^3 \text{x} \bar N(\text{x}) H_g (\text{x}) \equiv  (2\kappa)^{-1}\int_M d^3 \text{x}\bar  N(\text{x}) \frac{E^a_i E^b_j}{\sqrt{\det E}} \left[ {\epsilon^{ij}}_k F^k_{ab} +2(1+\gamma^2) K^i_{[a} K^j_{b]}\right](\text{x})
\\\\
G_g(\bar{\Lambda})\equiv \int_M d^3 \text{x} \bar{\Lambda}^i (\text{x})G_{g,i}(\text{x})\equiv (\gamma\kappa)^{-1} \int_M d^3 \text{x} \bar{\Lambda}^i \left( \partial_a E^a_i +{\epsilon_{ij}}^k A_a^j E^a_k \right)(\text{x})\rule{56pt}{0pt}
\\\\
M_g(\bar V)\equiv \int_M d^3 \text{x} \bar V^a(\text{x}) M_{g,a}(\text{x})\equiv (\gamma\kappa)^{-1} \int_M d^3 \text{x} \bar V^a(\text{x}) \left( E^b_i F^i_{ab} - \kappa(1-\gamma^2) K^i_a G_{g,i}\right)(\text{x})
\\
\end{split}
\end{equation}
The explicit forms of the matter terms depend on the matter fields, which we would specify later. 

The three constraints $(2.4)$ form a closed algebra with structure functionals independent of the explicit forms of the matter terms. Denoting $[ \bar\Lambda, \bar\Lambda']$ to be the $SU(2)$ commutator, and setting $[ \bar V, \bar V']^a\equiv \bar V^b \partial_b  \bar V'^a- \bar V'^b \partial_b  \bar V^a$, the algebra is given by
\begin{equation}
\begin{split}
\\
\{G(\bar{\Lambda}), G(\bar{\Lambda}')\}= 8\pi( G/c^3 ) \gamma G([\bar{\Lambda},\bar{\Lambda}']);\rule{20pt}{0pt}  \{G(\bar{\Lambda}), M(\bar V)\}= 8\pi( G/c^3 ) \gamma G(\mathcal{L}_{\bar V} \bar{\Lambda})\rule{90pt}{0pt}
\\\\
\{M(\bar V), M(\bar V')\}=  8\pi( G/c^3 ) \gamma M([\bar V,\bar V'])\rule{280pt}{0pt}
\\\\
\{G(\bar{\Lambda}), H(\bar N)\}= 0;\rule{20pt}{0pt} \{M(\bar V), H(\bar N)\}=  8\pi( G/c^3 ) \gamma H(\mathcal{L}_{\bar V}\bar N)\rule{175pt}{0pt}
\\\\
\{H(\bar N), H(\bar N')\}= 8\pi( G/c^3 )\gamma\left( M(\bar S)+G(\bar S^a A_a)\right)+ \frac{1-\gamma^2}{8\pi( G/c^3 )}\gamma G\left(\frac{[E^a\partial_a \bar N,E^b\partial_b \bar N']}{|\det E|}\right)\rule{60pt}{0pt}
\\
\\
\end{split}
\end{equation}
where $\mathcal{L}_{\bar V}$ denotes a Lie derivative and
\begin{equation}
\begin{split}
\bar S^a=( \bar N\partial_b \bar N'-\bar N'\partial_b \bar N) \frac{E^b_i E^{ai}}{|\det E|}\rule{80pt}{0pt}\\
\end{split}
\end{equation}

\subsection{ Knot Space}

 In QCD, the generalized electric flux and magnetic holonomy variables are powerful in capturing non-perturbative degrees of freedom of gluon fields. The Ashtekar formalism enables an analogous treatment of gravitational fields.
Gravitational holonomy and flux variables are signatures of loop quantum gravity \cite{intro,intro1,perez}, and they capture non-perturbative degrees of freedom of gravitational fields. The holonomy variable over an oriented path $\bar{e}\subset M$ (the bar indicates that $\bar{e}$ is embedded in the spatial manifold $M$) gives the parallel transport along the path by the connection fields $A^{i}_{b}(\text{x})$. The flux variable over an oriented surface $\bar{S}\subset M$ gives the flux of  ${E}^{a}_{i}(\text{x})$ through the surface. Explicitly, we have:
\begin{equation}
 {h}^{(j)}(\bar{e})^{\bar{k}}_{\bar{l}}[A]\equiv [\mathcal{P} \exp \int_{\bar{e}}d\bar{e}^{b} A^{i}_{b}(\text{x})\tau_{i}^{(j)}]^{\bar{k}}_{\bar{l}};\rule{10pt}{0pt}
 {F}_i(\bar{S})
\equiv \int_{\bar{S}}\hat{E}^{a}_{i}d\bar{S}_{a} 
\end{equation}
where $\mathcal{P}$ denotes path ordering along $\bar{e}$, and the $SU(2)$-valued gravitational holonomy is written in the spin $(j)$ matrix representation. The matter variables compatible with the gravitational flux and holonomy variables are the following  \cite{matter1,matter2,thiemann}. The gauge fields $\underbar{A}^{\text i}_a(\text{x})$ are described by the $\mathcal G$ holonomies $\text{h}^{(\text{i})}(\bar{e})^{\bar{\text i}}_{\bar{\text j}}$ in representations $(\text{i})$, and the $\underbar{E}_{\text i}^a(\text{x})$ fields are described by the flux variables ${\text F}_{\text i}(\bar{S})$. If $\bar v$ is a generic point in $M$, the $\xi^{\bar i}_{\bar{\text i}}(\text{x})$ fields are described by the irreducible tensors $\theta^{(d)}(\bar{v})$ obtained from their Grassmann monomials of degree $d$, and the spinor momenta $\pi_{\bar i}^{\bar{\text i}}(\text{x})$ are described by ${\eta}(\bar v)\equiv i{\theta}^{(d=1)}(\bar v)^{\dagger}$. The $\phi^{\text{j}}(\text{x})$ fields are described by $h^{(\text{k})}(\bar v) \equiv \exp(\phi^{\text{j}} (\bar v) {\tau^{(\text{k})}}_{\text{j}})$ called point holonomies in representations $(\text{k})$, and the momenta $P_{\text i}(\text{x})$  are described by ${p}_{\text i}(\bar v)$. This new set of gravitational and matter variables are collectively called \emph{loop variables}. The kinematical states of loop quantum gravity \cite{intro}\cite{intro1}\cite{perez} are called knot states, and they are functionals of the configuration Ashtekar variables $\{A^i_a(\text{x}),\underbar{A}^{\text i}_a(\text{x}),\xi^{\bar i}_{\bar{\text i}}(\text{x}),\phi _{\text i}(\text{x})\}$ via the corresponding loop variables $\{ {h}^{(j)}(\bar{e}),\text{h}^{(\text{i})}(\bar{e}), \theta^{(d)}(\bar{v}), h^{(\text{k})}(\bar v)\}$.

 A knot state is given by an $SU(2)\times \mathcal G$ invariant product of loop variables defined on a graph in $M$, so it solves the Gauss constraints. Further, the knot state is only sensitive to the embedding of the graph up to a spatial diffeomorphism $\mu\in \mathit{diff_M}$, so it also solves the momentum constraint. Define an embedded graph $\bar{\gamma}$ in $M$ to consist of $N_e$ smooth oriented paths $\{\bar{e}_i\}$, called edges, meeting at most at their end points $\{\bar{v}_n\}$, called nodes (the bars again indicate that $\bar{\gamma}$ is embedded in $M$). Carrying loop variables, an embedded colored graph $\bar \Gamma$ is defined by: 1) an embedded graph $\bar{\gamma}$; 2) an $SU(2)$ spin representation $j_{i}$ and a $\mathcal{G}$ group representation $\text{j}_i$ assigned to each edge;  3) generalized $SU(2)$ and $\mathcal{G}$ Clebsch-Gordan coefficients (intertwiners) $i_{n}$ and $\text{i}_n$, a point holonomy representation $\text{k}_n$ and a Grassmann monomial degree $d_n$ assigned to each node. The assignment of  $j_{i}$, $\text{j}_i$, $i_{n}$, $\text{i}_n$, $\text{k}_n$ and $d_n$ gives a $SU(2)\times \mathcal{G} $ scalar functional $S_{\bar{\Gamma}}$: 
\begin{equation}
\begin{split}
S_{\bar{\Gamma}}[A,\underbar{A}, \xi,\phi ]
\equiv Inv\left\{\bigotimes_n^{N_v} i_n \bigotimes_n^{N_v}\text{i}_n\bigotimes_n^{N_v}\theta^{(d_n)} (\bar{v}_n)\bigotimes_n^{N_v} h^{(\text{k}_n)}(\bar{v}_n) \bigotimes_i^{N_e}{h}^{(j_i)}(\bar{e}_i)  \bigotimes_i^{N_e}\text{h}^{(\text j_i)}(\bar{e}_i)\right\}[A,\underbar{A}, \xi,\phi]\\\\
\end{split}
\end{equation}
where $Inv\{ ...\}$ denotes the $SU(2) \otimes\mathcal{G}$ invariant contraction. 
 An element $\mu\in \mathit{diff_M}$ drags $\bar \Gamma$ to $\bar\Gamma'\equiv\mu\bar\Gamma$ and transforms $S_{\bar{\Gamma}}$ to $S_{\bar{\Gamma}}\hat\mu= S_{\bar\Gamma'}$. Note that for every $\bar\gamma$, there is a subgroup $\mathit{diff_{M,\bar{\gamma}}}$ of $\mathit{diff_M}$ that leaves the graph invariant, maintaining the \emph{set} of all the edges and their orientations. There is also a subgroup $\mathit{Tdiff_{M,\bar{\gamma}}}$ of $\mathit{diff_{M,\bar{\gamma}}}$ that acts trivially on $\bar \gamma$. Thus we have the graph symmetry group $G_{M,\bar{\gamma}}\equiv  \mathit{diff_{M,\bar{\gamma}}}/ \mathit{Tdiff_{M,\bar{\gamma}}}$. The group averaging operator $\hat{\mathbb P}_{\mathit{diff_M}}$ serves to erase the embedding information from $S_{\bar{\Gamma}}$, and it is defined as:
\begin{equation}
\begin{split}
S_{\bar{\Gamma}}\cdot\hat{\mathbb P}_{\mathit{diff_M}}\equiv S_{\bar{\Gamma}}\cdot\bigg[\frac{1}{N_{G_{M,\bar{\gamma}}}} \sum_{\mu \in G_{M,\bar{\gamma}}}\hat{\mu}\rule{3pt}{0pt} \cdot\rule{3pt}{0pt}\sum_{\mu'\in \mathit{diff_M}/\mathit{diff_{M,\bar{\gamma}}}}\hat\mu' \bigg]\equiv  s_{[\bar\Gamma]}
\end{split}
\end{equation}
where $N_{G_{M,\bar{\gamma}}}$ is the number of elements in $G_{M,\bar{\gamma}}$, and $[\bar\Gamma]$ is a colored graph obtained from the embedded colored graph $\bar\Gamma$ by erasing its exact embedding.\footnote{ Note that when $\bar\Gamma'=\bar\Gamma \hat\mu$, we have $[\bar\Gamma]=[\bar\Gamma']$.}
The result $s_{[\bar\Gamma]}$ is a knot state. Such a state is determined by a colored graph $[\bar\Gamma]$, and is $\mathit{diff_M}$ invariant since $s_{[\bar{\Gamma}]}\cdot \hat\mu=s_{[\mu\bar{\Gamma}]}= s_{[\bar{\Gamma}]}$.
The inner products between knot states are given by (generalized) Ashtekar-Lewandowski measure \cite{intro}\cite{intro1}\cite{perez}\cite{thiemann}, with which the set of all knot states $\{ \langle s_{[\bar\Gamma]}|\}$ gives an orthonormal basis:
\begin{equation}
\begin{split}
\langle s_{[\bar\Gamma']}|s_{[\bar\Gamma]}\rangle=\delta_{[\bar\Gamma'], [\bar\Gamma]}
\end{split}
\end{equation}
This basis spans knot space $K$, the $SU(2)\times \mathcal G$ and $\mathit{diff_M}$ invariant kinematical Hilbert space of loop quantum gravity.

Canonical quantization of loop variables leads to the operators of the form $\hat O(\bar{\Omega})$, where $\bar{\Omega}\subset M$ may be $\bar v$, $\bar e$, $\bar S$ or $\bar R$. However, $\hat O(\bar{\Omega})$ does not preserve $K$ since $\bar{\Omega}$ is not $\mathit{diff_M}$ invariant.
To respect $\mathit{diff_M}$ symmetry, we now replace $\bar{\Omega}$ by a dynamical object $\Omega$ that assigns $\bar\Omega( \bar\Gamma)\subset M$ to each $\bar\Gamma$, such that $\bar\Omega( \bar\Gamma)$ transforms together with $\bar\Gamma$ under $\mathit{diff_M}$ transformations. Setting $\{ \bar\Gamma_{rep}\}$ to contain exactly one representative from every $[\bar\Gamma]$, one can construct $\Omega$ by specifying $\bar\Omega( \bar\Gamma_{rep})$ for every $\bar\Gamma_{rep}\in\{ \bar\Gamma_{rep}\}$. More formally, each dynamical object $\Omega$ is a map $\Omega: \bar\Gamma \to \bar{\Omega}(\bar\Gamma)\subset M$ satisfying $ \bar \Omega(\mu\bar \Gamma)= \mu'\mu \bar \Omega(\bar \Gamma)$ for any $\mu \in \mathit{diff_M}$ and some $\mu' \in  \mathit{Tdiff_{M,{\mu\bar\gamma}}}$. Then, we can define $\mathit{diff_M}$ invariant operator $\hat O(\Omega) $ as:
\begin{equation}
\begin{split}
 s_{[\bar\Gamma]}\cdot \hat O(\Omega)\equiv S_{\bar\Gamma}\cdot\hat O(\bar{\Omega}(\bar\Gamma))\hat{\mathbb P}_{\mathit{diff_M}}\rule{10pt}{0pt}
\end{split}
\end{equation}
For example, one can construct $\mathit{diff_M}$ invariant flux and holonomy operators using a dynamical surface $S: \bar\Gamma \to \bar{S}(\bar\Gamma)\subset M$ and a dynamical path $e: \bar\Gamma \to \bar{e}(\bar\Gamma)\subset M$:
\begin{equation}
\begin{split}
 s_{[\bar\Gamma]}\cdot \hat{F}_i(S)\equiv S_{\bar\Gamma}\cdot\hat{F}_i(\bar{S}(\bar\Gamma))\hat{\mathbb P}_{\mathit{diff_M}};\rule{10pt}{0pt}
 s_{[\bar\Gamma]} \cdot \hat{h^{(j)}}({e})^{\bar{k}}_{\bar{l}}\equiv S_{\bar\Gamma}\cdot \hat{h^{(j)}}(\bar{e}(\bar\Gamma))^{\bar{k}}_{\bar{l}}\hat{\mathbb P}_{\mathit{diff_M}}
\end{split}
\end{equation}
The two operators are respectively differential and multiplicative operators acting on $s_{[\bar\Gamma]}$, and
their $SU(2)\times \mathcal G$ invariant products give gravitational operators in $K$.
Since $E^a_i (\text{x})$ determines the spatial metric, the spatial area and volume operators are made up of the flux operators \cite{intro,intro1,perez}. $\hat{F}_i(\bar S)$ as a differential operator acts on the gravitational sector of $S_{\bar\Gamma}$ in a way that satisfies the Leibniz rule. Specifically, when $\bar S$ intersects with $\bar\gamma$ only at its node $\bar v_1$, we have:
\begin{equation}
\begin{split}
\hat{F}_{i}(\bar{S})\cdot \bigotimes_n^{N_v}  i_n \bigotimes_i^{N_e} {h}^{(j_i)}(\bar{e}_i)
\equiv\sum_{\bar e_{i'}|_{\bar v_1 \in \bar e_{i'}}} \iota(\bar{S},\bar{e}_{i'})\iota(\bar{e}_{i'}, \bar v_1)\hat{J}_{i}(\bar e_{i'}) \cdot \bigotimes_n^{N_v}  i_n \bigotimes_i^{N_e} {h}^{(j_i)}(\bar{e}_i)\rule{100pt}{0pt}\\
\hat{J}_{i}(\bar e)\cdot\bigotimes_n^{N_v}  i_n \bigotimes_i^{N_e} {h}^{(j_i)}(\bar{e}_i)
\equiv i\hbar\kappa\gamma \sum_k \left [\delta_{\bar e, \bar{e}_k}\cdot{h}^{(j_k)}(\bar{e}_k)\tau^{(j_k)}_i-\delta_{\bar e, {\bar{e}_k}^{-1}}\cdot\tau^{(j_k)}_i{h}^{(j_k)}(\bar{e}_k)\right]\bigotimes_{n}^{N_v}  i_n \bigotimes_{i\neq k}^{N_e} {h}^{(j_i)}(\bar{e}_i)\\\\
\end{split}
\end{equation}
where $\iota(\bar{S},\bar{e}_{i'})$ is $+1$ or $-1$ when $\bar{e}_{i'}$ is above or below $\bar S$ in an infinitesimal neighborhood of $\bar v_1$, and is $0$ if otherwise; $\iota(\bar{e}_{i'}, \bar v_1)$ is $+1$ or $-1$ when $\bar v_1$ is the source or target of $\bar{e}_{i'}$. 

The area operator in loop quantum gravity is derived from regularizing the classical expression in terms of the flux variables. The resulting area operator of a dynamical surface ${S}$ is \cite{area}
\begin{equation}
s_{[\bar{\Gamma}]}\cdot\hat{A}_{{S}}\equiv S_{\bar{\Gamma}}\cdot\sqrt{\hat{F}_{i}(\bar{S}(\bar\Gamma))\hat{F}^{i}(\bar{S}(\bar\Gamma))}\hat{\mathbb P}_{\mathit{diff_M}}
\end{equation}
Remarkably, knot states are eigenstates of $\hat{A}_{{S}}$ for any dynamical surface ${S}$. Set $\bar{p}$ to be a point where $\bar{\gamma}$ intersects transversely with $\bar{S}(\bar\Gamma)$, and let $j_{i_p}$ represent the spin carried by the edge $\bar e_{i_p}$ containing the point $\bar{p}$. One can then easily check that 
\begin{equation}
\begin{split}
s_{[\bar{\Gamma}]}\cdot\hat{A}_{{S}}=8\pi l^2_p \gamma\sum_{\bar{p}} \sqrt{j_{i_p}(j_{i_p}+1)}s_{[\bar{\Gamma}]}
\end{split}
\end{equation}
where the Planck length $l_p$ is defined by $l^2_p\equiv\hbar (G/c^3)$. It is clear that the spectrum of area operators in $K$ is discretized, built from multiples of small quanta of order of $l^2_p$.

The volume operator in loop quantum gravity also results from the regularization of the classical expression using the flux variables, but with much more complicated technicalities. Also, there are a variety of different regularization approaches that lead to different operator forms \cite{volume,volume1}. Here we will follow the one introduced in \cite{volume1}. The volume operator given by this approach acts on $s_{[\bar{\Gamma}]}$ with a set of flux operators over dynamical surfaces intersecting $\bar\gamma$ only at the nodes. With the help of $(2.13)$, the action of the volume operator for a dynamical region $R$ boils down to \cite{volume1}:
\begin{equation}
\begin{split}
s_{[\bar{\Gamma}]}\hat{V}_{{R}}\equiv S_{\bar{\Gamma}} \sum_{\bar{v}_n\in \bar{R}(\bar\Gamma)} \sqrt{\big|\hat{q}_{\bar{v}_n} \big|}\hat{\mathbb P}_{\mathit{diff_M}}\rule{2pt}{0pt};\rule{5pt}{0pt}
\hat{q}_{\bar{v}_n} \equiv  \frac{1}{48}\sum_{\bar{v}_n\in\bar{e}_i,\bar{v}_n\in\bar{e}_j,\bar{v}_n\in\bar{e}_k}\text{sgn}(\bar{e}_{i},\bar{e}_{j},\bar{e}_{k}) \epsilon^{pqr}  \hat{J}_{p}(\bar{e}_{i})  \hat{J}_{q}(\bar{e}_{j}) \hat{J}_{r}(\bar{e}_{k})\rule{5pt}{0pt}
\end{split}
\end{equation}
By $(2.16)$ the volume operator for a region $\bar{R}$ acts only on the nodes of a knot state contained in ${R}$. Further, each node-wise operation on the spin network state is given by the sum of the triplets of the action $(2.13)$ on the edges meeting at the node. Therefore, the volume operator changes only the intertwiners of the nodes of $[\bar{\Gamma}]$. Consequently, there is a volume and area eigenbasis for knot space, consisting of states given by a definite $\bar{\gamma}$ and $j_i$.

Each member of the eigenbasis for the area and volume operators has the following interpretation:  a certain quantum of volume is assigned to each node of the state, and a certain quantum of area is assigned to each edge of the state.  Through this basis, $K$ provides a concrete description of quantum spatial geometry -- the networks carrying Planck sized units of areas and volumes, associated with their edges and nodes.

The remaining constraint to be imposed for a physical Hilbert space is the Hamiltonian constraint. Adhering to the polymer-like structure of knot states, the standard Hamiltonian constraint operator $\hat{H}(\bar N)^{LQG}\equiv\hat{H}^{LQG}_g(\bar N)+\hat{H}^{LQG}_m(\bar N)$ \cite{intro,perez,thiemann} is quantized from a regularized discrete expression approximating ${H}(\bar N)$. In the discrete expression, the curvature factors in ${H}(\bar N)$ are approximated by holonomies along a certain set of tiny loops. The quantization then leads to $\hat{H}(\bar N)^{LQG}$ that contains holonomy operators based on these loops. Therefore, the action of $\hat{H}(\bar N)^{LQG}$ on a knot state $s_{[\bar\Gamma]}$ involves a change in the graph topology that adds the set of tiny loops to $\bar\gamma$.
Moreover, since a non-constant Lagrangian constraint $\bar N$ is not $\mathit{diff_M}$ invariant, $\hat{H}^{LQG}(\bar N)$ does not preserve $K$ in general. With constant $\bar N$ the action of $\hat{H}^{LQG}(\bar N)$ preserves $K$, but it is intricate and changes the topology of the graphs. As a result, constructing a physical Hilbert space annihilated by $\hat{H}^{LQG}(\bar N)$ is a major challenge, which is currently tackled by both canonical approaches \cite{intro1,perez,group1,ave2} and the path integral formalisms \cite{perez,foam2,foam1}.

\section{The Model}

 In our model, we will modify $\hat{H}^{LQG}(\bar N)$ into a graph topology preserving operator $\hat H( N_p)$ defined on $K$. The greatly simplified setting will allow us to apply the group averaging method to construct the physical Hilbert space $\mathbb H$ of the model, based on certain concrete assumptions. A set of local Dirac observables in $\mathbb H$ will be obtained using the clock constructed from the $\phi$ field and spatial matter coordinates and frames from the $\{\psi\}$ sector.

\subsection{Modified Hamiltonian Constraint Operator and Physical Hilbert Space}

For each embedded graph $\bar\gamma$, we label each of its nodes with an integer $n$ and each of its edges connected to the node $n$ by an integer pair $(n,i)$ (the range of $i$ depends on $n$ in general).\footnote{Note that each edge has two labels since it contains two nodes.} For a given $\bar\gamma$, we denote its node $n$ by $\bar v_n^{\bar\gamma}$, and the oriented path starting from $\bar v_n^{\bar\gamma}$ and overlapping exactly with its edge $(n,i)$ by $\bar e_{n,i}^{\bar\gamma}$ (fig.1a). To each pair $(\bar e_{n,i}^{\bar\gamma},\bar e_{n,j}^{\bar\gamma})$ we assign a minimal oriented closed path $\bar e_{n,i,j}^{\bar\gamma}$ that lies in $\bar\gamma$, containing the outgoing path $\bar e_{n,i}^{\bar\gamma}$ and incoming path $(\bar e_{n,j}^{\bar\gamma })^{-1}$ (fig.1a). Using these labels, we define a set of dynamical nodes $\{ v_m \}$, satisfying $\bar v_m(\bar\Gamma )=\bar v_{n(m)}^{\bar\gamma}$ such that $n(m)$ is one-to-one. Corresponding to a given $\{v_m\}$, we define a set of dynamical paths $\{ e_{m,j}\}$ satisfying $\bar e_{m,j}(\bar\Gamma )=\bar e_{n(m),k(j)}^{\bar\gamma }$ and $\bar v_m(\bar\Gamma )=\bar v_{n(m)}^{\bar\gamma}$ such that $j(k)$ is one-to-one. The dynamical closed path $e_{m,i,j}$ is then determined by the outgoing $ e_{m,i}$ and incoming $ e^{-1}_{m,j}$ dynamical paths. We also define a set of dynamical spatial points $p\equiv\{p_k\}$, where $k$ ranges from $0$ to infinity, such that $\bar v_n^{\bar\gamma}=\bar p_k(\bar\Gamma )$ holds for exactly one $k$ for every $\bar\Gamma$ and $n$.\footnote{The spatial manifold $M$ contains uncountably many spatial points, but we use only countably infinite set $\{p_k\}$ in correspondence to the discrete structure of $K$. Also, the one-to-one correspondence between $\bar v_n^{\bar\gamma}$ and $\bar p_k(\bar\Gamma )$ is obvious had we taken $\{\bar p_k(\bar\Gamma)\}$ for any $\bar\Gamma$ to be the set of $all$ $points$ in $M$. Here we define $\{\bar p_k(\bar\Gamma)\}$ to be a countably infinite subset of $M$, while maintaining this natural condition.} Notice that there are infinitely many distinct sets of dynamical nodes, paths and spatial points satisfying the above. This ambiguity comes from the arbitrariness of identifying the nodes and edges between different knot states, due to the absence of a reference background in $K$.

\begin{figure}
\begin{center}
\includegraphics[angle=0,width=1.6in,clip=true]{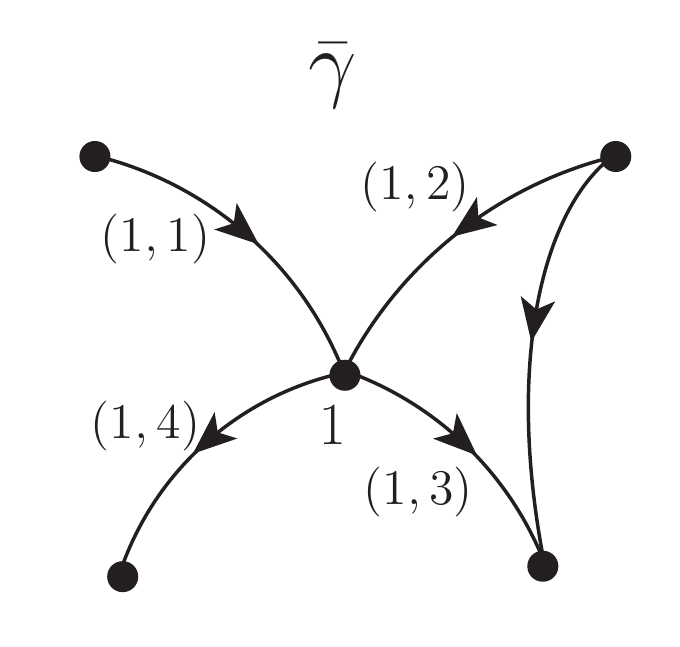}
\includegraphics[angle=0,width=4in,clip=true]{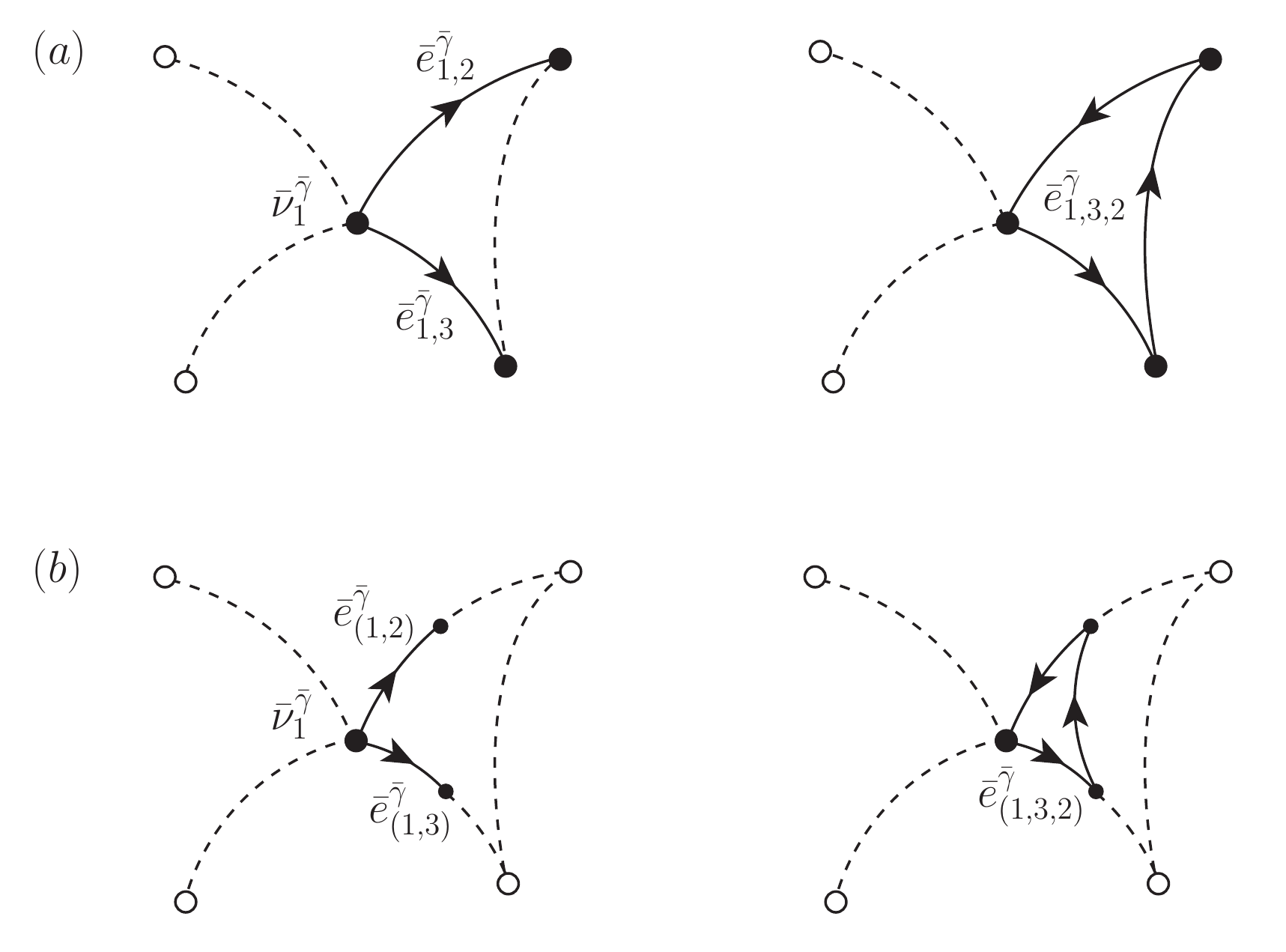}
\caption{The left figure depicts an embedded graph $\bar\gamma$, with one of its nodes labeled by $n=1$ and the four edges connected to this node labeled by $(1,j)$. The figures in $(a)$ demonstrate the corresponding definitions of $\bar v_n^{\bar\gamma}$, $\bar{e}^{\bar\gamma}_{n,j}$ and $\bar{e}^{\bar\gamma}_{n,i,j}$, while the figures in $(b)$ demonstrate the corresponding definitions of $\bar{e}^{\bar\gamma}_{(n,j)}$ and $\bar{e}^{\bar\gamma}_{(n,i,j)}$ that are used in $\hat{H}^{LQG}(\bar N)$.}
\end{center}
\end{figure}

The modification from $\hat{H}^{LQG}(\bar N)$ into $\hat H( N_p)$-- a crucial step for the model-- contains two elements \cite{lin1}. First, $\bar N$ is replaced by a $\mathit{diff_M}$ invariant lapse function $N_p(p_k)$, which is a function of a set of dynamical spatial points $p=\{p_k\}$. Recall that the embedded spatial point $\bar p_k(\bar\Gamma)\in M$ transforms under $\mathit{diff_M}$ just as $\bar\Gamma$ does. That means the classical counterpart of $N_p(p_k)$ is the field $N(\text x)$ that transforms as a scalar field. Therefore, in proper semi-classical limits, the operator $\hat H( N_p)$ is expected to approximate the $\mathit{diff_M}$ invariant ${H}_{g}(N)$ instead of ${H}_{g}(\bar N)$. Second, the tiny loops (fig.1b) that define the holonomy operators in $\hat{H}^{LQG}(\bar N)$ are replaced by the new set of closed paths (fig.1a) that are contained in the graphs of  knot states. By using the corresponding new set of holonomy operators, $\hat H( N_p)$ preserves the graph-topology of knot states. Since $\hat H( N_p)$ preserves both $\mathit{diff_M}$ symmetry and the graph topology, it is an operator in any subspace $K_{\mathcal{T}}\subset K$ with one specific graph topology $\mathcal{T}$. For simplicity, the model's kinematical Hilbert space is set to be $K_{\mathcal{T}_{torus}}\subset K$ with the graph topology of a lattice torus $\mathcal{T}_{torus}$, which has $N_v$ nodes and six edges connected to each of the nodes. Restricted to $K_{\mathcal{T}_{torus}}$, we have $1\leq n\leq N_v$ and $1\leq i\leq 6$ for $v_n$ and $ e_{n,i}$, and $\bar e_{n,i,j}^{\bar\gamma}$ will be a square closed path overlapping with exactly four edges in $\bar\gamma$

The modification applied to the gravitational term of $\hat{H}^{LQG}(\bar N)$ results to the new gravitational term $\hat H'_g(N_p)$, which acts on $\langle s_{[\bar\Gamma]}|\in K_{\mathcal{T}_{torus}}$ as (set $\hat{h} (\bar e)\equiv\hat{h}^{(1/2)}(\bar e)$):
\begin{equation}
\begin{split}
 s_{[\bar\Gamma]}\hat{H}'_g(N_p)
\equiv S_{\bar\Gamma}\bigg[\hat{H}'^{E}_{g(\bar\Gamma)}(N_p) +\frac{4(1+\gamma^2)}{8\kappa^4\gamma^7(i\hbar)^{5}}\sum_{ v_m}  N_p(  p_k|_{ \bar p_k(\bar\Gamma )=\bar v_m(\bar\Gamma)}) \sum_{i,j,k=1} \text{sgn}\left(\bar e_{m,i}(\bar\Gamma ),\bar e_{m,j}(\bar\Gamma ),\bar e_{m,k}(\bar\Gamma )\right)\rule{110pt}{0pt}\\ \hat{h}^{-1} (\bar e_{m,i}(\bar\Gamma) )_{\bar{i}}^{\bar{l}} \left[\hat{h}(\bar e_{m,i}(\bar\Gamma ))_{\bar{l}}^{\bar{j}},\left[\hat{H}^{E}_{g(\bar\Gamma)}(1),\hat{V}_{(\bar\Gamma)}\right] \right]\hat{h}^{-1}(\bar e_{m,j}(\bar\Gamma ))_{\bar{j}}^{\bar{p}}\left[\hat h(\bar e_{m,j}(\bar\Gamma ))_{\bar{p}}^{\bar{k}},\left[\hat{H}^{E}_{g(\bar\Gamma)}(1),\hat{V}_{(\bar\Gamma)}\right] \right]\rule{100pt}{0pt}\\\times\hat{h}^{-1} (\bar e_{m,k}(\bar\Gamma )_{\bar{k}}^{\bar{q}}\left[\hat{h}(\bar e_{m,k}(\bar\Gamma )_{\bar{q}}^{\bar{i}},\hat{V}_{(\bar\Gamma)}\right]\bigg]\hat{\mathbb P}_{diff}\rule{335pt}{0pt}
\\
\end{split}
\end{equation}
 where we have
\begin{equation}
\begin{split}
\\
\hat{H}'^{E}_{g(\bar\Gamma)}(N_p)
\equiv \frac{2}{16\kappa^2\gamma(i\hbar)}\sum_{ v_m}   N_p(  p_k|_{ \bar p_k(\bar\Gamma )=\bar v_m(\bar\Gamma)})  \sum_{i,j,k=1} \text{sgn}\left(\bar e_{m,i}(\bar\Gamma ),\bar e_{m,j}(\bar\Gamma ),\bar e_{m,k}(\bar\Gamma )\right)\rule{200pt}{0pt}\\
      \left(\hat{ h}(\bar e_{m,i,j}(\bar\Gamma)) -\hat{ h}^{-1}(\bar e_{m,i,j}(\bar\Gamma))
      \right) ^{\bar{i}}_{\bar{j}} \left(\hat{h}^{-1} (\bar e_{m,k}(\bar\Gamma ))\right)_{\bar{i}}^{\bar{l}}\cdot \left[\left(\hat{h}(\bar e_{m,k}(\bar\Gamma ))\right)_{\bar{l}}^{\bar{j}},\hat{V}_{(\bar\Gamma)} \right]\rule{185pt}{0pt}
\end{split}
  \end{equation}
Here, the total volume operator $\hat{V}_{(\bar\Gamma)}$ is defined as
\begin{equation}
\begin{split}\\
\hat{V}_{(\bar\Gamma)}
\equiv \sum_{ v_m}\bigg[\frac{1}{48}\sum_{i,j,k=1} \text{sgn}\left(\bar e_{m,i}(\bar\Gamma ),\bar e_{m,j}(\bar\Gamma ),\bar e_{m,k}(\bar\Gamma )\right) \epsilon^{pqr}\hat{J}_{p}(\bar e_{m,i}(\bar\Gamma ))\hat{J}_{q}(\bar e_{m,j}(\bar\Gamma ))  \hat{J}_{r}(\bar e_{m,k}(\bar\Gamma ))\bigg]^{\frac{1}{2}}\rule{155pt}{0pt}\\
\end{split}
\end{equation}
The modification applied to the matter term in $\hat{H}^{LQG}(\bar N)$ \cite{thiemann} results to the new matter term $\hat{H}_m(N_p)$, whose explicit form depends on the matter content. Similar to the pure gravitational term, it is also constructed from the loop operators in $K_{\mathcal{T}_{torus}}$, and acts on $\langle s_{[\bar\Gamma]}|\in K_{\mathcal{T}_{torus}}$ as:
\begin{equation}
\begin{split}
  s_{[\bar\Gamma]}\hat{H}'_{m}(N_p)\equiv  S_{\bar\Gamma}\hat{H}'_{m(\bar\Gamma)}(N_p)\hat{\mathbb P}_{diff}\rule{350pt}{0pt}
\\
\hat{H}'_{m(\bar\Gamma)}(N_p)\equiv\sum_{ v_n} N_p(  p_k|_{ \bar p_k(\bar\Gamma )=\bar v_n(\bar\Gamma)})
\hat{H}'^{v_n}_{m}\bigg(\hat{J}_{i}(\bar e_{n,i}(\bar\Gamma )),\hat{h}(\bar e_{n,i}(\bar\Gamma ))_{\bar{l}}^{\bar{j}},\hat{h}( \bar e_{n,i,j}(\bar\Gamma))_{\bar{l}}^{\bar{j}},\hat{\text{J}}_{\text i}(\bar e_{n,i}(\bar\Gamma )),\rule{82pt}{0pt}\\ \hat{\text{h}}^{(\text j)}(\bar e_{n,i}(\bar\Gamma ))^{\bar{\text i}}_{\bar{\text j}}, \hat{\text{h}}^{(\text j)}(\bar e_{n,i,j}(\bar\Gamma))^{\bar{\text i}}_{\bar{\text j}}\hat {\theta} (\bar{ v}_n(\bar\Gamma ))^{\bar i}_{\bar{\text i}}, \hat {\theta} (\bar v_{n,i}(\bar\Gamma ))^{\bar i}_{\bar{\text i}},\hat{\eta} (\bar{ v}_n(\bar\Gamma ))_{\bar i}^{\bar{\text i}},\hat{\eta} (\bar v_{n,i}(\bar\Gamma ))_{\bar i}^{\bar{\text i}}, \hat{h}^{(\text{i})}(\bar{ v}_n(\bar\Gamma ))^{\bar{\text i}}_{\bar{\text j}},\rule{40pt}{0pt}\\\hat{h}^{(\text{i})}(\bar v_{n,i}(\bar\Gamma ))^{\bar{\text i}}_{\bar{\text j}},\hat{p}_{\text i}(\bar{ v}_m(\bar\Gamma )),\hat{p}_{\text i}(\bar v_{n,i}(\bar\Gamma )) \bigg)\rule{263pt}{0pt}\\
\end{split}
\end{equation}
where $v_{n,i}$ denotes the end node of $e_{n,i}$.

Finally, the Hamiltonian constraint operator for our model is given by the self-adjoint sum of the gravitational and matter terms: 
\begin{equation}
\hat H( N_p)\equiv\frac{1}{2}(\hat H'_g(N_p)+\hat H'_m(N_p))+\frac{1}{2}(\hat H'_g(N_p)+\hat H'_m(N_p))^{\dagger}
\nonumber
\end{equation}

In order to compare the model with loop quantum cosmology, we will choose an identical physical setting. In this setting, gravitational fields are fully coupled to the chargeless, massless real scalar field $\phi$, while all other matter fields with small back-reaction are collectively named as $m'$.  The classical constraints in this case are:
\begin{equation}
\begin{split}
\\
H(\bar N) 
= H_{g}(\bar N)+ H_{\phi}(\bar N)+ H_{\{\psi\}}(\bar N)\rule{360pt}{0pt}
\\
\equiv (2\kappa)^{-1}\int_M d^3 \text{x}\bar  N(\text{x}) \frac{E^a_i E^b_j}{\sqrt{\det E}} \left[ {\epsilon^{ij}}_k F^k_{ab}
+2(1-\gamma^2) K^i_{[a} K^j_{b]}\right](\text{x})\rule{216pt}{0pt}\\
+(2\kappa)^{-1}\int_M d^3 \text{x}\bar N(\text{x}) \frac{1}{\sqrt{\det E}}\left[\delta^{ij}E^a_i E^b_j(\partial_a   \phi)\partial_b \phi+ (2\kappa)^2 P^2\right](\text{x})
+ H_{\{\psi\}}(\bar N)\rule{139pt}{0pt}
\\
\equiv H_{g,\phi}(\bar N)+ H_{\{\psi\}}(\bar N)\rule{400pt}{0pt}
\\\\
G(\bar{\Lambda})
=G_g(\bar{\Lambda})+G_{\phi}(\bar{\Lambda})+G_{\{\psi\}}(\bar{\Lambda})\rule{370pt}{0pt}
\\
= (\gamma\kappa)^{-1} \int_M d^3 \text{x} \bar{\Lambda}^i \left( \partial_a E^a_i +{\epsilon_{ij}}^k A_a^j E^a_k \right)(\text{x})+ 0+ G_{\{\psi\}}(\bar{\Lambda})\rule{233pt}{0pt}
\\
\equiv G_{g,\phi}(\bar{\Lambda})+G_{\{\psi\}}(\bar{\Lambda})\rule{407pt}{0pt}
\\\\
M(\bar V)
=M_g(\bar V)+M_{\phi}(\bar V)+M_{\{\psi\}}(\bar V)\rule{362pt}{0pt}
\\
=(\gamma\kappa)^{-1} \int_M d^3 \text{x} \bar V^a(\text{x}) \left( E^b_i F^i_{ab} - \kappa(1-\gamma^2) K^i_a G_{g,i}\right)(\text{x})\rule{252pt}{0pt}\\
+\int_M d^3 \text{x}\bar V^a(\text{x}) P\partial_a \phi(\text{x}) +M_{\{\psi\}}(\bar V)\rule{328pt}{0pt}
\\
\equiv M_{g,\phi}(\bar V)+M_{\{\psi\}}(\bar V) \rule{403pt}{0pt}
\end{split}
\end{equation}
Note that the functionals $ H_{g,\phi}(\bar N)$, $G_{g,\phi}(\bar{\Lambda})$ and $M_{g,\phi}(\bar V)$ are the constraints for the subsystem of only gravitational and $\phi$ fields, and themselves satisfy the algebra $(2.5)$.

 The loop operators for $\phi$ field are the point holonomy opreators $\hat{h}^{(\lambda)}(\bar v) \equiv \exp(i\lambda\hat{\phi} (\bar v))$ with arbitrary real $\lambda$, and the conjugate momenta operators $\hat{p}(\bar v)$. Their non-vanishing commutators are given by
\begin{equation}
[\hat{h}^{(\lambda)}(\bar v) ,\hat{p}(\bar v')]=i\hbar \delta_{\bar v, \bar v'}(i\lambda)\hat{h}^{(\lambda)}(\bar v)
\end{equation}
Corresponding to $(3.5)$, we have $\hat H'_m(N_p) =\hat{ H}'_{{\phi}}(N_p)+\hat{ H}'_{\{\psi\}}(N_p)$.  
The modified operator $\hat{ H}'_{{\phi}}(N_p)$ formed by quantizing $H_{\phi}( N)$ is given by \cite{thiemann}:
\begin{equation}
\begin{split}
 s_{[\bar\Gamma]}\hat{H}'_{{\phi}}(N_p)
\equiv S_{\bar\Gamma}\frac{-81^2}{64^2\cdot48^2\cdot 2\kappa}(i\hbar\kappa\gamma)^{-4}\frac{1}{\lambda^2} \sum_{v_m} N_p(  p_k|_{ \bar p_k(\bar\Gamma)=\bar v_m(\bar\Gamma)})\hat D^q(\bar v_m(\bar\Gamma))\hat D_q(\bar v_m(\bar\Gamma))\hat{\mathbb P}_{diff}\rule{243pt}{0pt}
   \\
+ S_{\bar\Gamma}\frac{64^2\cdot 2\kappa}{27^2\cdot48^2}(i\hbar\kappa\gamma)^{-6}\sum_{v_m} N_p(  p_k|_{ \bar p_k(\bar\Gamma)=\bar v_m(\bar\Gamma)})\hat{p}(\bar v_m(\bar\Gamma))\hat{p}(\bar v_m(\bar\Gamma))\bigg\{\sum_{i,j,k} \text{sgn}\left(\bar e_{m,i}(\bar\Gamma ),\bar e_{m,j}(\bar\Gamma ),\bar e_{m,k}(\bar\Gamma )\right)\rule{197pt}{0pt}\\
\left(\hat{h}^{-1} (\bar e_{m,i}(\bar\Gamma ))\right)_{\bar{J}}^{\bar{I}}\left[\left(\hat{h}(\bar e_{m,i}(\bar\Gamma ))\right)_{\bar{K}}^{\bar{J}},\hat{V}^{\frac{1}{2}}(\bar v_m(\bar\Gamma))\right]
\left(\hat{h}^{-1} (\bar e_{m,j}(\bar\Gamma ))\right)_{\bar{L}}^{\bar{K}}\left[\left(\hat{h}(\bar e_{m,j}(\bar\Gamma ))\right)_{\bar{M}}^{\bar{L}},\hat{V}^{\frac{1}{2}}(\bar v_m(\bar\Gamma))\right]\rule{199pt}{0pt}\\\times
 \left(\hat{h}^{-1} (\bar e_{m,k}(\bar\Gamma ))\right)_{\bar{N}}^{\bar{M}}\left[\left(\hat{h}(\bar e_{m,k}(\bar\Gamma ))\right)_{\bar{I}}^{\bar{N}},\hat{ V}^{\frac{1}{2}}(\bar v_m(\bar\Gamma))\right]\bigg\}^2\hat{\mathbb P}_{diff}\rule{373pt}{0pt}
\end{split}
\end{equation}
where 
\begin{equation}
\begin{split}
\hat D^q(\bar v_m(\bar\Gamma))
\equiv\sum_{i,j,k} \text{sgn}\left(\bar e_{m,i}(\bar\Gamma ),\bar e_{m,j}(\bar\Gamma ),\bar e_{m,k}(\bar\Gamma )\right)
 {\hat h}^{(\lambda_0)}(\bar v_m(\bar\Gamma))^{-1}\left[{\hat h^{(\lambda_0)}}(\bar e_{m,i}(\bar\Gamma)_{(1)})- {\hat h^{(\lambda_0)}}(\bar v_m(\bar\Gamma))\right]\rule{75pt}{0pt}\\ \times(\tau^q)_{\bar{m}}^{\bar{n}}\left(\hat{h}^{-1} (\bar e_{m,j}(\bar\Gamma))\right)_{\bar{k}}^{\bar{m}}\left[\left(\hat{h}(\bar e_{m,j}(\bar\Gamma))\right)_{\bar{l}}^{\bar{k}},\hat{ V}^{\frac{3}{4}}(\bar v_m(\bar\Gamma))\right]\rule{200pt}{0pt} \\
\times\left(\hat{h}^{-1}(\bar e_{m,k}(\bar\Gamma))\right)_{\bar{p}}^{\bar{l}}\left[\left(\hat{h}(\bar e_{m,k}(\bar\Gamma))\right)_{\bar{n}}^{\bar{p}},\hat{ V}^{\frac{3}{4}}(\bar v_m(\bar\Gamma))\right]\rule{228pt}{0pt}\\
\end{split}
\end{equation}
with $\lambda_0$ a small real number whose exact value is unimportant in our context.
The modified Hamiltonian constraint operator $\hat H( N_p)$ in our setting is then given by:
\begin{equation}
\begin{split}
\\
\hat H( N_p)
\equiv\frac{1}{2}(\hat H'_g(N_p)+\hat H'_m(N_p))+\frac{1}{2}(\hat H'_g(N_p)+\hat H_m'(N_p))^{\dagger}
\equiv \hat H_{g,\phi}(N_p)+\frac{1}{2}(\hat H'_{\{\psi\}}+\hat H'^{\dagger}_{\{\psi\}})(N_p)
\end{split}
\end{equation}

Recall that our kinematical Hilbert space $K_{\mathcal{T}_{torus}}$ is already $SU(2)\times \mathcal G$ and $\mathit{diff_M}$ invariant. To obtain the physical Hilbert space of the model, we still need to impose the remaining symmetry generated by $\{\exp(i\hat{H}(N_p))\}$ with arbitrary $N_p$ based on arbitrary $p$. These unitary operators form a faithful representation of a group $G$, that is
\begin{equation}
\left\{\prod_{k=1}^{\infty}\exp(i\hat{H}(N_{p_k}))\right\}_{\sim K_{\mathcal{T}_{torus}}}  \equiv \left\{ \hat{U}(g)\right\}_{g\in G}
\end{equation}
where $N_{p_k}$ is an arbitrary lapse function based on an arbitrary $p_k$, and $\sim K_{\mathcal{T}_{torus}}$ means that we identify two expressions if they give the same operator in $K_{\mathcal{T}_{torus}}$.
The physical Hilbert space of the model is constructed using group averaging procedure under the assumptions: (1) the existence of the left and right invariant measure $dg$ for $G$; (2) the operator $\hat{\mathbb{P}}$ defined by 
\begin{equation}
\hat{\mathbb{P}} \equiv \int_G dg \hat{U}(g) 
\end{equation}
maps $\langle \psi|\in K_{\mathcal{T}_{torus}}$ into $\hat{\mathbb P}|\psi\rangle\in K_{\mathcal{T}_{torus}}^*$. The two conditions hold in minisuperspace models \cite{ave1,marolf1,marolf2}, but remain to be proven for the model.
Under these assumptions, the inner product between any two states $|\Psi_1\rangle=\hat{\mathbb{P}}|\psi_1\rangle$ and $| \Psi_2\rangle=\hat{\mathbb{P}}|\psi_2\rangle$ may be defined as
\begin{equation}
\langle\Psi_1|\Psi_2\rangle\equiv\langle\psi_1|\Psi_2\rangle= \langle\psi_1|\hat{\mathbb{P}}|\psi_2\rangle \\
\end{equation}

The physical Hilbert space of the model $\mathbb H \subset  K_{\mathcal{T}_{torus}}^*$ is the space spanned by $\{\hat{\mathbb P}|s_{[\bar\Gamma]}\rangle\}$. By construction, $\mathbb H$ is invariant under the action of $\{\exp(i\hat{H}(N_p))\}$ with arbitrary $N_{p}$, and satisfies the modified Hamiltonian constraint. Each element in $\mathbb H$ is a solution to the quantized  Einstein equations, and therefore is a quantum state of spacetime with the matter fields.

\subsection{Local Dirac Observables}

The model obtains its local Dirac observables using clocks, spatial coordinates and frames given by the matter fields. For an arbitrary set of dynamical nodes $\{v_m\}$, one may construct a set of self-adjoint, commuting matter operators consisting of scalar operators $\{ \hat\phi^{0}(v_m),\hat\phi^{1}(v_m), \hat\phi^{2}(v_m), \hat\phi^{3}(v_m)\}$ diagonalized by the knot state basis of $K_{\mathcal{T}_{torus}}$, current operators $\{\hat{V^{i}_{I}}(v_m), \hat{U}^{\bar{i}}_{\bar{I}}(v_m)\}$ and conjugate current operators $\{\hat{\bar{V}}^{I}_{i}(v_m),\hat{\bar{U}}^{\bar{I}}_{\bar{i}} (v_m)\}$ ($I=1,2, 3$ for the vector currents; $\bar{I}=1,2$ for the spinor currents). The operator triplet $(\hat{\phi}^{1}(v_m), \hat{\phi}^{2}(v_m), \hat{\phi}^{3}(v_m))\equiv \hat{\Phi} (v_m)$ will serve as spatial coordinate operators, $\{\hat{V^{i}_{I}}(v_m), \hat{U}^{\bar{i}}_{\bar{I}}(v_m)\}$ and $\{\hat{\bar{V}}^{I}_{i}(v_m), \hat{\bar{U}}^{\bar{I}}_{\bar{i}} (v_m)\}$ will be spatial frame operators, and $\hat\phi^{0}(v_m)$ will be the clock operator. In our setting, the cosmological clock field $\hat\phi^{0}(v_m)$ is chosen to be:
\begin{equation}
\begin{split}
\hat\phi^{0}(v_m)\equiv(2i\lambda_0)^{-1}[{\hat h}^{(\lambda_0)}( v_m)- {\hat h}^{(-\lambda_0)}( v_m)]
\end{split}
\end{equation}
provided by the $\phi$ field.

Classical gravitational fields in the Ashtekar formalism are $SU(2)$ tensors. In the model, the $SU(2)$ invariant components of the fields are described relative to the matter spatial frames. Explicitly, for any $ s_{[\bar\Gamma]}\in K_{\mathcal{T}_{torus}}$ we define
\begin{equation}
\begin{split}
s_{[\bar\Gamma]}\cdot\hat{J}( e_{n, j})_{I} \equiv S_{\bar\Gamma} \cdot\hat{V}(\bar v_n(\bar \Gamma))^{i}_{I}\hat{J}(\bar e_{n, j}(\bar \Gamma))_{i}\hat{\mathbb P}_{diff}\rule{55pt}{0pt}\\
 s_{[\bar\Gamma]}\cdot\hat{h}(e_{n,k})^{\bar{I}}_{\bar{J}} \equiv  S_{\bar\Gamma}\cdot\hat{\bar{U}}(\bar v_{n,k}(\bar \Gamma))^{\bar{I}}_{\bar{i}} \hat{h}(\bar e_{n,k}(\bar \Gamma))^{\bar{i}}_{\bar{j}}\hat{U}(\bar v_n(\bar \Gamma))^{\bar{j}}_{\bar{J}}\hat{\mathbb P}_{diff}
\end{split}
\end{equation}
In canonical general relativity with the matter spatial coordinate field $\Phi(\text x)$, we can obtain a $\mathit{diff_M}$ invariant and spatially local variable $O(X)$ by integrating $\det(\partial_a \Phi(\text x))\delta(\Phi(\text x)-X) O(\text x)$ over $M$. Analogously, the model uses a normalized Gaussian distribution ${\delta}^{\epsilon}$ with finite width $\epsilon$ and defines:
\begin{equation}
\begin{split}
s_{[\bar\Gamma]}\hat{O}(X) \equiv  S_{\bar\Gamma}\sum_{n}\det(\Delta \hat{\Phi}(\bar v_n(\bar\Gamma))) \hat{\delta}^{\epsilon}( \hat{\Phi}(\bar v_n(\bar\Gamma))- X)\hat{O}(\bar v_n(\bar\Gamma))  \hat{\mathbb P}_{diff}\rule{146pt}{0pt}\\
s_{[\bar\Gamma]}\hat{O'}(e_{X,\Delta X})\equiv S_{\bar\Gamma}\sum_{n,i} \det(\Delta \hat{\Phi}(\bar v_n(\bar\Gamma))) \det(\Delta \hat{\Phi}(\bar v_{n,i}(\bar\Gamma)))\hat{\delta}^{\epsilon}(\hat{\Phi}(\bar v_n(\bar\Gamma))-X) \hat{\delta}^{\epsilon}(\hat{\Phi}(\bar v_{n,i}(\bar\Gamma))-X-\Delta X)\\\times\hat{O}(\bar e_{n,i}(\bar\Gamma))\hat{\mathbb P}_{diff}\rule{315pt}{0pt}\\
\end{split}
\end{equation}
 for any $ s_{[\bar\Gamma]}\in K_{\mathcal{T}_{torus}}$,
where the coordinate volume element operators are given by:
\begin{equation}
\begin{split}
\Delta \hat{\Phi}_{\bar e_{n,i}(\bar \Gamma)}\equiv  \big[\hat{\Phi}(\bar v_{n,i}(\bar\Gamma)) - \hat{\Phi}(\bar v_n(\bar \Gamma))\big]\rule{290pt}{0pt}
\\
 S_{\bar\Gamma}\det(\Delta \hat{\Phi}(\bar v_n(\bar \Gamma)))
\equiv  S_{\bar\Gamma}\sum_{(i,j,k)}\text{sgn}\left(\bar e_{n,i}(\bar \Gamma),\bar e_{n,j}(\bar \Gamma),\bar e_{n,k}(\bar \Gamma)\right) \Delta \hat{\Phi}_{\bar e_{n,i}(\bar \Gamma)}\cdot \left(\Delta \hat{\Phi}_{\bar e_{n,j}(\bar \Gamma)} \times\Delta \hat{\Phi}_{\bar e_{n,k}(\bar \Gamma)}\right)
\nonumber
\end{split}
\end{equation}
 It is crucial that the spatially local operators obtained in $(3.7)$ do not depend on the choices of the dummy variables $\{v_m\}$ and $\{e_{n,i}\}$. The $\mathit{diff_M}$ invariant operators are localized by referring to the spatial matter coordinates.
 
To specify time using the clock field $\phi^{0}$, we construct an operator $\hat{\Pi}_T$ that maps a spacetime state $|\Psi\rangle\in \mathbb H$ to the spatial slice state $\hat{\Pi}_T|\Psi\rangle\in K^*_{\mathcal{T}_{torus}}$ on which $\phi^{0}=T$. With $\hat{\omega}(\bar v)$ denoting an operator acting on $\bar v$, the operator $\hat{\Pi}_T$ is defined by:
\begin{equation}
\begin{split}
\hat{\nu}_{\omega} (\bar v_n(\bar\Gamma)) \equiv \frac{ i}{\hbar}\left[\hat{\omega}(\bar v_n(\bar\Gamma)), \hat{H}_{(\bar \Gamma)} (1) \right]\rule{114pt}{0pt}
\\
 s_{[\bar\Gamma]}\hat{\Pi}_T\equiv S_{\bar\Gamma} sym\left\{\prod_n \hat{\nu}_{\phi^0} (\bar v_n(\bar\Gamma)) \hat{\delta}^{\epsilon}(\hat{\phi}^0(\bar v_n(\bar\Gamma))- T)\right\}\hat{\mathbb P}_{diff}\rule{3pt}{0pt}
\end{split}
\end{equation}
where the symmetrization $sym\{...\}$  in the ordering of $n$ is applied. 
The resulting local Dirac observables in $\mathbb H$  are given by
\begin{equation}
\hat{O}(X, T)\equiv\hat{\mathbb{P}}\hat{O}(X)  \hat{\Pi}_T\rule{2pt}{0pt};\rule{4pt}{0pt}\hat{O'}(e_{X,\Delta X},T)\equiv\hat{\mathbb{P}}\hat{O'}(e_{X,\Delta X})  \hat{\Pi}_T
\end{equation}
In the following, $\omega$ in $(3.16)$ will serve as an index running over the matter coordinates and frames $\{\phi^0,\rule{2pt}{0pt}\Phi, \rule{3pt}{0pt}f\cdot V_I\rule{1pt}{0pt},\rule{3pt}{0pt} \bar f\cdot\bar V^I\rule{0pt}{0pt} ,\rule{3pt}{0pt}g\cdot U_{\bar I}\rule{1pt}{0pt}, \rule{3pt}{0pt}\bar g \cdot\bar U^{\bar I}\}$ contracted with the non-zero $SU(2)$ test functions $\{f^i,\bar f_i, g^{\bar i}, \bar g_{\bar i}\}$. The variables ${\nu}_{\omega}$ thus contain information about the momenta of the reference matter fields.

Finally, applying $(3.14)$, $(3.15)$ and $(3.17)$ to the fields of concern we obtain the relevant localized observables. In the following, we will give the dynamics of $\hat{J}(e_{X, \Delta X}, T)_{I}$, $\hat{h}(e_{X, \Delta X}, T)^{\bar{I}}_{\bar{J}}$, $\hat p(X,T)$, ${\hat h}^{(\lambda)}(X,T)$, in a certain ``physical gauge" of matter coordinates and frames that is fixed by $\hat{\nu}_{\omega}(X, T)$.

\subsection{Conditions on Matter Coordinates and Frames}
Our next step is to choose a proper physical state $|\Psi\rangle \in \mathbb H$ to derive semi-classical limits from the model.
 In order to give sensible descriptions, the localized observables must refer to well-behaved matter coordinates and frames. So there are conditions to be imposed on the matter sector of $|\Psi\rangle$, which provides the matter coordinates and frames. 

For the clock, we require any two spatial slices of $|\Psi\rangle$ with $\phi^{0}=T_1$ and $\phi^{0}=T_2$ to be related by causal dynamics, so that either one of them can be used to reconstruct the spacetime. In the model, this condition is imposed as:
\begin{equation}
\hat{ \mathbb{P}}\hat{\Pi}_{T_1}|\Psi\rangle+O(\hbar) = \hat{\mathbb{P}}\hat{\Pi}_{T_2}|\Psi\rangle+O(\hbar)=|\Psi\rangle
\end{equation}
When $(3.18)$ is satisfied, we have:
\begin{equation}
\begin{split}
\hat{\mathbb{P}}\hat{O}_1 (X) \hat{O}_2 (X) \hat{\Pi}_{T_1} |\Psi\rangle\rule{63pt}{0pt}\\
=\hat{\mathbb{P}}\hat{O}_1 (X) \hat{O}_2 (X) \hat{\Pi}_{T_1} \hat{\mathbb{P}} \hat{\Pi}_{T_1} |\Psi\rangle+O(\hbar)\\ 
=\hat{\mathbb{P}}\hat{O}_1 (X)  \hat{\Pi}_{T_1} \hat{\mathbb{P}} \hat{O}_2 (X) \hat{\Pi}_{T_1} |\Psi\rangle+O(\hbar)\\
= \hat{O}_1 (X,T_1) \hat{O}_2 (X, T_1)  |\Psi\rangle+O(\hbar)\rule{18pt}{0pt}
\end{split}
\end{equation}
More generally, denoting by $\varphi'(O_1, O_2, ....,O_N)$ a function of $N$ linear operators in a Hilbert space, we have
\begin{equation}
\begin{split}
\hat{\mathbb{P}} \varphi'(\hat{O}_1 (X), \hat{O}_2 (X),...., \hat{O}_N (X)) \hat{\Pi}_{T_1} |\Psi\rangle\rule{60pt}{0pt}\\
=  \varphi'(\hat{O}_1 (X, T_1), \hat{O}_2 (X, T_1),...., \hat{O}_N (X, T_1)) |\Psi\rangle+O(\hbar)
\end{split}
\end{equation}

On each spatial slice $\hat{\Pi}_{T_1}|\Psi\rangle$, we will also impose spatial coordinate conditions to properly describe gravitational and $\phi$ fields. Denote an arbitrary function of the gravitational and $\phi$ field operators based on $\{ e_{n,i}\}$ and $\{v_n\}$ as:
\begin{equation}
\hat\varphi\equiv \varphi( \hat{J}(e_{n,i})_{i} ,\hat{h}(e_{n,i})^{\bar{i}}_{\bar{j}},\hat{h}^{\dagger}(e_{n,i})^{\bar{i}}_{\bar{j}},\hat p(v_n),{\hat h}^{(\lambda)}(v_n), {\hat h}^{(\lambda)\dagger}( v_n))
\nonumber
\end{equation}
We require that there exist $\{v^*_n\}$ and $\{e^*_{n,i}\}$ such that:
\begin{equation}
\begin{split}
\varphi( \hat{J}(e^*_{n,i})_{i} ,\hat{h}(e^*_{n,i})^{\bar{i}}_{\bar{j}},\hat{h}^{\dagger}(e^*_{n,i})^{\bar{i}}_{\bar{j}},\hat p(v^*_n),{\hat h}^{(\lambda)}(v^*_n), {\hat h}^{(\lambda)\dagger}( v^*_n))\hat{\Pi}_{T_1}|\Psi\rangle\rule{136pt}{0pt}\\
= \varphi( \hat{J}(e_{X_n, \Delta X_{n,i}})_{I} ,\hat{h}(e_{X_n, \Delta X_{n,i}})^{\bar{I}}_{\bar{J}},\hat{h}^{\dagger}(e_{X_n, \Delta X_{n,i}})^{\bar{I}}_{\bar{J}},\hat p(X_n),{\hat h}^{(\lambda)}(X_n), {\hat h}^{(\lambda)\dagger}( X_n))\hat{\Pi}_{T_1}|\Psi\rangle+O(\hbar)
\end{split}
\end{equation}
with a certain set of values $\{X_n\}$ and $\{\Delta X_{m,j}\}$ satisfying $\{X_m+\Delta X_{m,j}\}=\{X_n\}$. Here, $\{v^*_ n\}$ represents the physical nodes at clock time $T_1$ for an observer using the spatial matter coordinates $\Phi=(\phi^1,\phi^2,\phi^3)$. Naturally, there is also a set of physical spatial points $p^*\equiv \{p^*_m\}$ agreeing with $\{v^*_n\}$, such that $ \bar p^*_m(\bar\Gamma)=\bar v^*_m(\bar\Gamma)$ for $m \le N_v$. Condition $(3.21)$ says that each physical node acquires a matter spatial coordinate value, and that the matter frames are physically orthonormal to each other at each physical node.
The spatial coordinate condition can now be imposed on the map $( v^*_n, e^*_{n,i}) \to (X_n, \Delta X_{n,i})$. Analogous to the coordinate maps on a torus manifold, the map should appear smooth in large scales at most of the physical nodes except at some $\{v^*_{n_b} \}\subset \{ v^*_{n}\}$ where coordinate singularities occur. We demand $|\Delta X_{n,i}| \leq d$ for $v^*_n \not\in \{ v^*_{n_b}\}$, where $d$ is small enough that the map appears continuous at large scales. Also, for any $m$ the set $\{\Delta X_{m,i}\leq d\}$ should define a parallelepiped in $\mathbb{R}^3$ up to an error of $O(d)$ so the map appears smooth at large scales. Notice that once $(3.11)$ is satisfied, the coordinate conditions can be achieved easily through a redefinition of coordinates  $\hat\Phi \to \hat\Phi'(\hat\Phi)$.

Using a lapse function $ N^{\small \mathcal{N}}_{p^*}$ satisfying $N^{\small \mathcal N}_{p^*}(p^*_m) = \mathcal{N}(X_{m})$ with an arbitrary function $\mathcal{N}(X)$, we can apply $(3.21)$ to $\hat{H}_{g,\phi}( N^{\small \mathcal{N}}_{p^*})$:
\begin{equation}
\begin{split}
\hat{H}_{g,\phi}( N^{\small \mathcal{N}}_{p^*})\hat{\Pi}_{T_1}|\Psi\rangle
\equiv\varphi( N^{\small \mathcal{N}}_{p^*}(p^*_n),\hat{J}(e^*_{n,i})_{i} ,\hat{h}(e^*_{n,i})^{\bar{i}}_{\bar{j}},\hat{h}^{\dagger}(e^*_{n,i})^{\bar{i}}_{\bar{j}},\hat p(v^*_n),{\hat h}^{(\lambda)}(v^*_n), {\hat h}^{(\lambda)\dagger}( v^*_n))\hat{\Pi}_{T_1}|\Psi\rangle\rule{35pt}{0pt}\\
= \hat{\mathcal{H}}_{g,\phi}(\mathcal N)\hat{\Pi}_{T_1}|\Psi\rangle+O(\hbar)\rule{276pt}{0pt}\\\\
\hat{\mathcal{H}}_{g,\phi}(\mathcal N)\equiv \varphi(\mathcal{N}(X_n),\hat{J}(e_{X_n, \Delta X_{n,i}})_{I} ,\hat{h}(e_{X_n, \Delta X_{n,i}})^{\bar{I}}_{\bar{J}},\hat{h}^{\dagger}(e_{X_n, \Delta X_{n,i}})^{\bar{I}}_{\bar{J}},\hat p(X_n),{\hat h}^{(\lambda)}(X_n), {\hat h}^{(\lambda)\dagger}( X_n))  \rule{24pt}{0pt}\\
\equiv\frac{1}{2}[ (\hat{\mathcal{H}}'_{g}(\mathcal N) +\hat{\mathcal{H}}'_{\phi}(\mathcal N))+(\hat{\mathcal{H}}'_{g}(\mathcal N) +\hat{\mathcal{H}}'_{\phi}(\mathcal N))^\dagger] \rule{208pt}{0pt}
\end{split}
\end{equation}
Copying the form of $\hat{H}'_{g}( N^{\small \mathcal{N}}_{p^*})$ shown in $(3.1)$, $\hat{\mathcal{H}}'_g(\mathcal N)$ is given by:
 \begin{equation}
\begin{split}
 \hat{\mathcal H}'_{g}(\mathcal{N})
\equiv \hat{\mathcal H}'^{E}_{g}(\mathcal{N}) - 2(1+\gamma^2)\frac{2}{8\kappa^4\gamma^7(i\hbar)^{5}} \sum_{X}\mathcal{N}(X)\sum_{\Delta X, \Delta Y, \Delta Z} \text{sgn}\left(e_{X, \Delta X},e_{X, \Delta Y},e_{X, \Delta Z}\right)\rule{110pt}{0pt}\\\times \left(\hat{h}^{-1} (e_{X,  \Delta X})\right)_{\bar{I}}^{\bar{L}} \left[\left(\hat{h}(e_{X,  \Delta X})\right)_{\bar{L}}^{\bar{J}},\left[\hat{\mathcal H}'^{E}_{g}(1),\hat{\mathcal V}\right] \right]\left(\hat{h}^{-1} (e_{X, \Delta Y})\right)_{\bar{J}}^{\bar{P}}\left[\left(\hat{h}(e_{X, \Delta Y})\right)_{\bar{P}}^{\bar{K}},\left[\hat{\mathcal H}'^{E}_{g}(1),\hat{\mathcal V}\right] \right]\rule{34pt}{0pt} \\\times \left(\hat{h}^{-1} (e_{X, \Delta Z})\right)_{\bar{K}}^{\bar{Q}}\left[\left(\hat{h}(e_{X, \Delta Z})\right)_{\bar{Q}}^{\bar{I}},\hat{\mathcal V}\right] \rule{300pt}{0pt}\\\\
\end{split}
\end{equation}
where $\hat{\mathcal H}'^{E}_{g}(\mathcal N)$ is
  \begin{equation}
\begin{split}
 \hat{\mathcal H}'^{E}_{g}(\mathcal{N})
\equiv\frac{2}{16\kappa^2\gamma(i\hbar)}\sum_{X}\mathcal{N}(X)\sum_{\Delta X, \Delta Y, \Delta Z} \text{sgn}\left(e_{X, \Delta X},e_{X, \Delta Y},e_{X, \Delta Z}\right)
       \left(   \hat{ h}(e_{X,  \Delta X, \Delta Y}) -\hat{ h}^{-1}(e_{X,  \Delta X, \Delta Y})
      \right) ^{\bar{I}}_{\bar{J}} \\ \times\left(\hat{h}^{-1} (e_{X, \Delta Z})\right)_{\bar{I}}^{\bar{L}}\cdot \left[\left(\hat{h}(e_{X, \Delta Z})\right)_{\bar{L}}^{\bar{J}}, \hat{\mathcal V} \right]\rule{239pt}{0pt}\\\\
\end {split}
\end{equation}
and
\begin{equation}
\begin{split}
 \hat{\mathcal V}= \sum_{X} \sum_{ \Delta X, \Delta Y, \Delta Z} \left[\frac{1}{48}\text{sgn}\left(e_{X, \Delta X},e_{X, \Delta Y},e_{X, \Delta Z}\right) \hat{\epsilon}^{PQR}\hat{J}(e_{X, \Delta X})_{P} \hat{J}(e_{X, \Delta Y})_{Q} \hat{J}(e_{X, \Delta Z})_{R}\right]^{\frac{1}{2}}\\\\
\end{split}
\end{equation} 
Copying the form of $\hat{H}'_{\phi}( N^{\small \mathcal{N}}_{p^*})$ shown in $(3.7)$, $\hat{\mathcal{H}}'_\phi(\mathcal N)$ is given by:
\begin{equation}
\begin{split}
\hat{\mathcal H}'_{{\phi}}(\mathcal{N})
\equiv\frac{-81^2}{64^2\cdot48^2\cdot 2\kappa}(i\hbar\kappa\gamma)^{-4} \frac{1}{\lambda^2} \sum_{X} \mathcal N(X)\hat D^Q(X)\hat D_Q(X)\rule{300pt}{0pt} \\
+\frac{64^2\cdot 2\kappa}{27^2\cdot48^2}(i\hbar\kappa\gamma)^{-6} \sum_{X}\mathcal{N}(X) \hat{p}(X)\hat{p}(X)\bigg\{\sum_{\Delta X, \Delta Y, \Delta Z} \text{sgn}\left(e_{X, \Delta X},e_{X, \Delta Y},e_{X, \Delta Z}\right)\rule{161pt}{0pt} \\
\times\left(\hat{h}^{-1} (e_{X, \Delta Y})\right)_{\bar{J}}^{\bar{I}}\left[\left(\hat{h}(e_{X, \Delta Y})\right)_{\bar{K}}^{\bar{J}},\hat{\mathcal V}^{\frac{1}{2}}(X)\right]
\left(\hat{h}^{-1} (e_{X, \Delta Y})\right)_{\bar{L}}^{\bar{K}}\left[\left(\hat{h}(e_{X, \Delta Y})\right)_{\bar{M}}^{\bar{L}},\hat{\mathcal V}^{\frac{1}{2}}(X)\right]\rule{125pt}{0pt}\\\
\times \left(\hat{h}^{-1} (e_{X, \Delta Z})\right)_{\bar{N}}^{\bar{M}}\left[\left(\hat{h}(e_{X, \Delta Z})\right)_{\bar{I}}^{\bar{N}},\hat{\mathcal V}^{\frac{1}{2}}(X)\right]\bigg\}^2\rule{307pt}{0pt}\\\\
\end{split}
\end{equation}
where 
\begin{equation}
\begin{split}
\hat D^Q(X)
\equiv\sum_{\Delta X, \Delta Y, \Delta Z} \text{sgn}\left(e_{X, \Delta X},e_{X, \Delta Y},e_{X, \Delta Z}\right)
 {\hat h}^{(\lambda)}(X)^{-1}\left[{\hat h^{(\lambda)}}(X+\Delta X)- {\hat h^{(\lambda)}}(X)\right]\rule{75pt}{0pt}\\ \times(\tau^Q)_{\bar{M}}^{\bar{N}}\left(\hat{h}^{-1} (e_{X, \Delta Y})\right)_{\bar{K}}^{\bar{M}}\left[\left(\hat{h}((e_{X, \Delta Y})\right)_{\bar{L}}^{\bar{K}},\hat{ V}^{\frac{3}{4}}(X)\right]\rule{185pt}{0pt} \\
\times\left(\hat{h}^{-1}(e_{X, \Delta Z})\right)_{\bar{P}}^{\bar{L}}\left[\left(\hat{h}(e_{X, \Delta Z})\right)_{\bar{N}}^{\bar{P}},\hat{ V}^{\frac{3}{4}}(X)\right]\rule{225pt}{0pt}\\
\end{split}
\end{equation}

Further, the well-behaved matter coordinates and frames also lead to simple algebraic relations between the spatially local operators acting on $\hat\varphi\hat{\Pi}_{T_1}|\Psi\rangle\equiv |\varphi\rangle$. From now on we set a variable $(X, \Delta X)$ to take values in $\{(X_n, \Delta X_{n,i})\}$. Also, define $\hat{\alpha}(e_{Y, \Delta Y})$ as the operator localized from $\hat{\alpha}(e_{n,i})$ which is diagonalized by the basis $\{\langle s_{[\bar \Gamma]}| \in K_{\mathcal{T}_{torus}}\}$. The state $\langle s_{[\bar \Gamma]}|$ has eigenvalue $+1$ if the embedded edge of $\bar \Gamma$ overlapping with $\bar e_{n,i}(\bar \Gamma)$ has the same orientation as $\bar e_{n,i}(\bar \Gamma)$ and has eigenvalue $-1$ if otherwise. The algebraic relations between the spatially local operators are the following:
\begin{equation}
\begin{split}
\hat{J}^\dagger(e_{X, \Delta X})_{I}|\varphi\rangle=\hat{J}(e_{X, \Delta X})_{I} |\varphi\rangle+O(\hbar)\rule{209pt}{0pt}\\\\
\left[\hat{J}(e_{X, \Delta X})_{I}, \hat{h}(e_{Y, \Delta Y})^{\bar{I}}_{\bar{J}}\right]|\varphi\rangle= \delta_{X,Y} \delta_{\Delta X,\Delta Y}i l_p^2\gamma  ({\tau _I})^{\bar{K}}_{\bar{J}} \hat{h}(e_{Y, \Delta Y})^{\bar{I}}_{\bar{K}}|\varphi\rangle\rule{87pt}{0pt}\\ - \delta_{X,Y+ \Delta Y} \delta_{-\Delta X,\Delta Y}i l_p^2 \gamma({\tau _I})^{\bar{I}}_{\bar{K}} \hat{h}(e_{Y, \Delta Y})^{\bar{K}}_{\bar{J}}|\varphi\rangle+O( l_p^2 \hbar)\\\\
\left[\hat{J}(e_{X, \Delta X})_{I}, \hat{h}^{\dagger}(e_{Y, \Delta Y})^{\bar{I}}_{\bar{J}}\right]|\varphi\rangle= \delta_{X,Y} \delta_{\Delta X,\Delta Y}i l_p^2 \gamma ({\tau^* _I})^{\bar{K}}_{\bar{J}} \hat{h}^{\dagger}(e_{Y, \Delta Y})^{\bar{I}}_{\bar{K}}|\varphi\rangle\rule{79pt}{0pt}\\ - \delta_{X,Y+ \Delta Y} \delta_{-\Delta X,\Delta Y}i l_p^2\gamma ({\tau^*_I})^{\bar{I}}_{\bar{K}} \hat{h}^{\dagger}(e_{Y, \Delta Y})^{\bar{K}}_{\bar{J}}|\varphi\rangle+O( l_p^2 \hbar)\\\\
\left[\hat{h}(e_{X, \Delta X})^{\bar{K}}_{\bar{L}}, \hat{h}(e_{Y, \Delta Y})^{\bar{I}}_{\bar{J}}\right] |\varphi\rangle= 0+O( l_p^2  \hbar)\rule{203pt}{0pt}\\\\
\left[\hat{h}(e_{X, \Delta X})^{\bar{K}}_{\bar{L}}, \hat{h}^\dagger(e_{Y, \Delta Y})^{\bar{I}}_{\bar{J}}\right] |\varphi\rangle= 0+O( l_p^2  \hbar)\rule{200pt}{0pt}\\\\
\left[\hat{J}(e_{X, \Delta X})_{I}, \hat{J}(e_{Y, \Delta Y})_{J}\right]|\varphi\rangle= \delta_{X,Y} \delta_{\Delta X,\Delta Y}i l_p^2\gamma{\epsilon_{IJ}}^K \hat{J}(e_{Y, \Delta Y})_{K} \hat{\alpha}(e_{Y, \Delta Y})|\varphi\rangle+O( l_p^2  \hbar)\\\\
\hat{p}(X) |\varphi\rangle=\hat{p}^{\dagger}(X)  |\varphi\rangle+O(\hbar)\rule{272pt}{0pt}\\\\
\left[ \hat{p}(X),\hat{h}^{(\lambda)}(Y)\right]|\varphi\rangle= \delta_{X,Y}i\hbar (i\lambda) \hat{h}^{(\lambda)}(Y)|\varphi\rangle+O(\hbar^2)\rule{152pt}{0pt}\\\\
\left[ \hat{p}(X),\hat{h}^{(\lambda)\dagger}(Y)\right]|\varphi\rangle= -\delta_{X,Y}i\hbar (i\lambda) \hat{h}^{(\lambda)\dagger}(Y)|\varphi\rangle+O(\hbar^2)\rule{135pt}{0pt}\\\\
\left[ \hat{h}^{(\lambda)}(X),\hat{h}^{(\lambda')}(Y)\right]|\varphi\rangle= 0+O(\hbar^2)\rule{233pt}{0pt}\\\\
\left[ \hat{h}^{(\lambda)}(X),\hat{h}^{(\lambda')\dagger}(Y)\right]|\varphi\rangle= 0+O(\hbar^2)\rule{231pt}{0pt}\\\\
\left[ \hat{p}(X),\hat{p}(Y)\right]|\varphi\rangle= 0+O(\hbar^2)\rule{265pt}{0pt}\\
\end{split}
\end{equation}
 This conditional algebra enables further calculations after the approximations $(3.21)$ are made.

Lastly, we also require $|\Psi\rangle$ to give the momenta of the matter coordinates and frames:
\begin{equation}
\hat{\nu}_{\omega}(X, T_1)|\Psi\rangle={\nu}_{\omega}(X, T_1)|\Psi\rangle+O(\hbar)
\end{equation}
The matter coordinates and frames specify a gauge for the gravitational sector. Locally at the moment $T_1$, this gauge is characterized by $(3.21)$ and $(3.29)$. 
For this paper, we require the simple gauge condition:\footnote{Note that the vaules of ${\nu}_{\omega}(X, T)$ can be adjusted by clock-time dependent redefinitions of the matter coordinates and frames}
\begin{equation}
{\nu}_{\phi^0}(X, T)\neq 0\rule{2pt}{0pt};\rule{4pt}{0pt}{\nu}_{\Phi}(X, T)={\nu}_{f\cdot V_I}(X, T)={\nu}_{\bar f\cdot\bar V^I}(X, T)={\nu}_{g\cdot U_{\bar I}}(X, T)={\nu}_{\bar g \cdot\bar U^{\bar I}}(X, T)=0
\end{equation}
which gives a comoving frame in the cosmological setting that will follow later.

\subsection{ Coherent States and Emergent Fields }

We have required the state $|\Psi\rangle$ to satisfy the quantum coordinate conditions $(3.21)$ and $(3.29)$, such that the set of local observables $\{\hat{J}(e_{X, \Delta X}, T)_{I},\hat{h}(e_{X, \Delta X}, T)^{\bar{I}}_{\bar{J}}\}$ and $\{ {\hat h}^{(\lambda)}(X, T) ,\hat{p}(X,T)\}$ would give meaningful descriptions around the moment $T_1$. Since our goal is to obtain semi-classical limits, we impose coherence conditions on the gravitational and $\phi$ field sector of $|\Psi\rangle$ as:
\begin{equation}
\begin{split}
\hat{J}(e_{X, \Delta X}, T_1)_{I}|\Psi\rangle=\left(\langle\Psi|\hat{J}(e_{X, \Delta X}, T_1)_{I}|\Psi\rangle\right) |\Psi\rangle +O(l_p^2)\\\\
\hat{h}(e_{X, \Delta X}, T_1)^{\bar{I}}_{\bar{J}}|\Psi\rangle=\left(\langle\Psi|\hat{h}(e_{X, \Delta X}, T_1)^{\bar{I}}_{\bar{J}}|\Psi\rangle\right) |\Psi\rangle +O(l_p^2)\\\\
{\hat h}^{(\lambda)}(X, T_1)|\Psi\rangle=\left(\langle\Psi|{\hat h}^{(\lambda)}(X, T_1)|\Psi\rangle\right) |\Psi\rangle +O(\hbar)\rule{30pt}{0pt}\\\\
\hat{p}(X,T_1)|\Psi\rangle=\left(\langle\Psi|\hat{p}(X,T_1)|\Psi\rangle\right) |\Psi\rangle +O(\hbar)\rule{60pt}{0pt}\\\\
\end{split}
\end{equation} 
Since the clock $\phi^0$ is built from $\phi$ field, the conditions imply:
\begin{equation}
\hat{\nu}_{\phi^0}(X, T_{1})|\Psi\rangle=\left(\langle\Psi|\hat{\nu}_{\phi^0}(X, T_{1})|\Psi\rangle\right)|\Psi\rangle +O(\hbar)\rule{20pt}{0pt}\\
\end{equation}
which specifies the value in $(3.30)$.
Additionally, we also want the expectation values to appear continuous in terms of the spatial coordinates for a semi-classical state. Therefore, for any two $\bar e^*_{n,i}(\bar\Gamma)$ and $\bar e^*_{m,j}(\bar\Gamma)$ that share a common node and form a smooth path, we will impose (recall that $|\Delta X_{n,i}| \leq d$ for $v^*_n \not\in \{ v^*_{n_b}\}$):
\begin{equation}
\begin{split}
\langle\Psi|\hat{J}(e_{X_n, \Delta X_i}, T_1)_{I}|\Psi\rangle= \langle\Psi|\hat{J}(e_{X_m, \Delta X_j}, T_1)_{I}|\Psi\rangle +O(d)\\\\
\langle\Psi|\hat{h}(e_{X_n, \Delta X_i}, T_1)^{\bar{I}}_{\bar{J}}|\Psi\rangle=\langle\Psi|\hat{h}(e_{X_m, \Delta X_j}, T_1)^{\bar{I}}_{\bar{J}}|\Psi\rangle+O(d)\\\\
\langle\Psi|{\hat h}^{(\lambda)}(X_n, T_1)|\Psi\rangle=\langle\Psi|{\hat h}^{(\lambda)}(X_m, T_1)|\Psi\rangle +O(d)\rule{35pt}{0pt}\\\\
\langle\Psi|\hat{p}(X_n,T_1)|\Psi\rangle=\langle\Psi|\hat{p}(X_m,T_1)|\Psi\rangle +O(d)\rule{60pt}{0pt}\\\\
\end{split}
\end{equation} 
Because of the algebraic relation $(3.28)$, we expect the gravitational and $\phi$ field sector's solutions to $(3.32)$ and $(3.33)$ to exist. These conditions say that $|\Psi \rangle$ has sharply defined, approximately continuous values for the local observables at clock time $T_1$. Therefore, $|\Psi\rangle$ is expected to be semi-classical around that moment. 

To make contact with classical general relativity, the model maps the expectation values in $(3.31)$ to the classical field values, using the matter coordinates as a common reference. 
For an explicit example, we will first pick a simple spatial matter coordinate system. Recall that the lattice torus $\mathcal{T}_{torus}$ can be constructed by identifying the opposite boundary faces of a lattice rectangular prism $\bar I_{\mathbb{Z}}^3 \subset \mathbb{R}^3$, which consists of the vertices $\{\bar V_n=X_n\}$ and links $\{\bar l_i\}$ (the bars indicate the embedding in $\mathbb{R}^3$). Such construction naturally gives a coordinate map $ v^*_n\to \bar V_n=X_n$. Also, the approximated spatial coordinate space is naturally a rectangle region $\bar I^3\supset\bar I_{\mathbb{Z}}^3 $ inside of $\mathbb{R}^3$.
 
Once matter spatial coordinates are chosen, our model identifies every $ e^*_{n.i}$ disjoint from $\{ v^*_{n_b}\}$ with an embedded path in $\bar I^3 $, under the guidance of the matter coordinate values. In our case, $ e^*_{n.i}$ is identified with the oriented path $\bar{e}_{X_n, \Delta X_{n,i}}$ that goes from the vertex $X_n$ to the vertex $X_n+\Delta X_{n,i} $, which overlaps exactly with the link $\bar l_{i'}$ connecting the two vertices. Subsequently, we choose a cell decomposition dual to $\bar I_{\mathbb{Z}}^3$, dividing $\bar I^3 $ into a set of cells $\{\bar c_{X_n }\}$ that are parallelepipeds up to errors of $O(d)$. Each cell $\bar c_{X_n }$ uniquely contains a vertex $X_n$, and the boundaries of $\{\bar c_{X_n }\}$ consist of a set of faces $\{\bar s_i\}$ whose each element $\bar s_{i'}$ intersects transversely with a unique link $\bar l_{i'}$. Suppose $\bar l_{i'}$ links $X_n$ and $X_n+\Delta X_{n,i} $, we denote $\bar{S}_{X_n, \Delta X_{n,i}}\subset \bar I^3$ the oriented surface overlapping with $\bar s_{i'}$ and having the same orientation as $\bar{e}_{X_n, \Delta X_{n,i}}$.
Then, model uses a fitting algorithm that maps the expectation values $\{\langle \hat{J}(e_{X, \Delta X}, T)_{I}\rangle, \langle\hat{h}(e_{X, \Delta X}, T)^{\bar{I}}_{\bar{J}}\rangle\}$ and $\{\langle{\hat h}^{(\lambda)}(X, T) \rangle,\langle\hat{p}(X,T)\rangle\}$ to the values of the smooth fields $\{ E^{a}_{I}( X, T ), A^{J}_{b}( X, T)\}$ and $\{ {\phi}(X, T), {P}(X,T)\}$ defined in $\bar I^{3}$. The fitting algorithm  is required to obey the following rules:\footnote{ Note that such an algorithm is guaranteed to exist, since we are fitting the smooth fields with infinite degrees of freedom to the finitely many data points given by the expectation values of the local observables.}
\begin{equation}
\begin{split}
  \int_{\bar{S}_{X, \Delta X}}E^{a}_{I}(T)ds_{a}  \equiv \langle \hat{J}(e_{X, \Delta X}, T )_I \rangle\rule{35pt}{0pt}\\\\
\mathcal{P}\exp[ \int_{\bar{e}_{X, \Delta X}} A^{J}_{b}(T)(\tau_{J})de^{b}]^{\bar{K}}_{\bar{L}} \equiv \langle \hat{h}(e_{X, \Delta X}, T)\rangle ^{\bar{K}}_{\bar{L}}\rule{32pt}{0pt}
\\\\
\int_{\bar c_{X }}{P}(X',T) dX'\equiv \langle\hat{p}(X,T)\rangle\rule{60pt}{0pt}
\\\\
\exp (i\lambda{\phi}(X, T))\equiv \langle{\hat h}^{(\lambda)}(X, T) \rangle\rule{50pt}{0pt}
\\\\
\end{split}
\end{equation}
In the rest of the paper, we will use the spatial coordinates with $\Delta X_{m,i}\in\{(\pm d, 0,0), (0,\pm d,0), (0,0,\pm d)\}$ for $v^*_m \not\in \{ v^*_{n_b}\}$, which makes $\bar c_{X_n }$ a right cubical cell, and $\bar{S}_{X, \Delta X}$ a square oriented surface.
The choice of $\bar I_{\mathbb{Z}}^3 \subset \mathbb{R}^3$, the cell decomposition $\{\bar s_i\}$, and the fitting algorithm described above are restricted but non-unique. However, any choice satisfying the restrictions gives a valid correspondence between $|\Psi\rangle$ and the emergent fields.

\section{Physics in Homogeneous, Isotropic and Spatially Flat Sector}

The paper \cite{lin1} has shown that the fields $E^{a}_{I}( X, T )$ and $A^{J}_{b}( X, T)$ determined by $|\Psi\rangle$ reproduce vacuum general relativity up to quantum gravitational corrections, when matter back reactions can be ignored. In the following a similar calculation will be done, assuming $|\Psi\rangle$ gives homogeneous, isotropic and spatially flat emergent fields $\{ E^{a}_{I}( X, T_1 ), A^{J}_{b}( X, T_1)\}$ and $\{ {\phi}(X, T_1), {P}(X,T_1)\}$ at a late initial time $T_1$. In comparison with \cite{lin1} the calculation will be done with two extensions: 1) while the matter back reactions from $\{\psi\}$ will still be ignored, the full interaction between gravitational and $\phi$ fields will be considered; 2) the quantum gravitational corrections will be evaluated to $O(\hbar^0)$. The calculation will give the $O(\hbar^0)$ effective equations governing the fields $\{ E^{a}_{I}( X, T ), A^{J}_{b}( X, T)\}$ and $\{ {\phi}(X, T), {P}(X,T)\}$, which will be compared with the $O(\hbar^0)$ effective equations in different models of loop quantum cosmology.

\subsection{ Emergent Constraints and Diffeomorphism Algebra}
 For notational simplicity, we will denote the collections of the spatially local gravitational, $\phi$ field and $\alpha$ operators as $\{\hat{J},\hat{h},\hat{p},{\hat h}^{(\lambda)}, \hat\alpha\}$. One can evaluate the n-fold commutator by first applying $(3.28)$:
\begin{equation}
\begin{split}
\langle\Psi|\hat{\mathbb P}\left(\frac{i}{\hbar}\right)^{n-1}\left[\hat{\mathcal H}_{g,\phi}(\mathcal{N}_n),....\left[\hat{\mathcal H}_{g,\phi}(\mathcal{N}_3),\left[\hat{\mathcal H}_{g,\phi}(\mathcal{N}_2), \hat{\mathcal H}_{g,\phi}(\mathcal{N}_1)\right]\right]...\right]\hat{\Pi}_{T_1}|\Psi\rangle\rule{20pt}{0pt}\\
=\langle\Psi|\hat{\mathbb P}\Phi_n(\hat{J},\hat{h},{\hat h}^{(\lambda)}, \hat{p},\hat\alpha, \mathcal{N}_i )\hat\Pi_{T_1}|\Psi\rangle+O(\hbar)\rule{161pt}{0pt}
\end{split}
\end{equation}
where $\Phi_n$ is given by carrying out all $n$ commutators. Then we use $(3.19)$, $(3.31)$ and $(3.34)$ to obtain:
\begin{equation}
\begin{split}
\langle\Psi|\hat{\mathbb P}\left(\frac{i}{\hbar}\right)^{n-1}\left[\hat{\mathcal H}_{g,\phi}(\mathcal{N}_n),....\left[\hat{\mathcal H}_{g,\phi}(\mathcal{N}_3),\left[\hat{\mathcal H}_{g,\phi}(\mathcal{N}_2), \hat{\mathcal H}_{g,\phi}(\mathcal{N}_1)\right]\right]...\right]\hat{\Pi}_{T_1}|\Psi\rangle\rule{70pt}{0pt}
\\
=\langle\Psi|\Phi_n(\hat{J}(T_1),\hat{h}(T_1),{\hat h}^{(\lambda)}(T_1), \hat{p}(T_1),\hat\alpha(T_1), \mathcal{N
}_i )|\Psi\rangle +O(\hbar)\rule{137pt}{0pt}
\\
=\Phi_n(\langle\hat{J}(T_1)\rangle,\langle\hat{h}(T_1)\rangle,\langle{\hat h}^{(\lambda)}(T_1)\rangle, \langle\hat{p}(T_1)\rangle,\langle\hat\alpha(T_1)\rangle, \mathcal{N}_i )+O(\hbar)\rule{126pt}{0pt}
\\
=\left\{{H}_{g,\phi}(\bar{N}_n),....\left\{{H}_{g,\phi}(\bar N_3),\left\{{H}_{g,\phi}(\bar{N}_2),{H}_{g,\phi}(\bar{N}_1)\right\}\right\}...\right\}\big|_{E_{I}^{a}(T_1), A_{b}^{J}(T_1),\phi( T_1), {P}(T_1), \bar{N}_i=\mathcal{N}_i}\rule{15pt}{0pt}
\\
+O(\hbar)+O(d^4)\rule{329pt}{0pt}
\\\\
\end{split}
\end{equation}
 Recall that ${H}_{g,\phi}(\mathcal{N}_i)$ is the Hamiltonian constraint for the subsystem of only gravitational and $\phi$ fields. The  $\{\psi\}$ sector serves as a background for this subsystem. Since the spatial coordinates are given by the $\{\psi\}$ sector, the lapse functions $\mathcal{N}_i (X)$ are Lagrangian multipliers \emph{for the subsystem}.  According to $(2.5)$, the Poisson brackets, to all orders in $n$, with arbitrary $\mathcal{N}_i$ reproduce the full (off-shell) diffeomorphism algebra between $ H_{g,\phi}(\bar N)$, $G_{g,\phi}(\bar{\Lambda})$ and $M_{g,\phi}(\bar V)$ in the semi-classical limit of $|\Psi\rangle$, up to corrections of $O(\hbar)+O(d^4)$.

Moreover, the symmetry $\hat H(N_p) |\Psi\rangle=0$ implies that:
\begin{equation}
\begin{split}
\langle\Psi|\hat{\mathbb P}\left(\frac{i}{\hbar}\right)^{n-1}\left[\hat{H}(N^{\mathcal{N}_n}_{p^*}),....\left[\hat{H}(N^{\mathcal{N}_3}_{p^*}),\left[\hat{H}(N^{\mathcal{N}_2}_{p^*}), \hat{H}(N^{\mathcal{N}_1}_{p^*})\right]\right]...\right]\hat{\Pi}_{T_1}|\Psi\rangle=0
\end{split}
\end{equation}
Separating the contributions involving $\hat{ H}'_{\{\psi\}}$ or $\hat{ H}'^{\dagger}_{\{\psi\}}$ and denoting them as $\epsilon_{\{\psi\}}$, we have:
\begin{equation}
\begin{split}
0=\langle\Psi|\hat{\mathbb P}\left(\frac{i}{\hbar}\right)^{n-1}\left[\hat{H}(N^{\mathcal{N}_n}_{p^*}),....\left[\hat{H}(N^{\mathcal{N}_3}_{p^*}),\left[\hat{H}(N^{\mathcal{N}_2}_{p^*}), \hat{H}(N^{\mathcal{N}_1}_{p^*})\right]\right]...\right]\hat{\Pi}_{T_1}|\Psi\rangle\rule{178pt}{0pt}\\
=\langle\Psi|\hat{\mathbb P}\left(\frac{i}{\hbar}\right)^{n-1}\left[\hat{ H}_{g,\phi}(N^{\mathcal{N}_n}_{p^*}),....\left[\hat{H}_{g,\phi}(N^{\mathcal{N}_3}_{p^*}),\left[\hat{ H}_{g,\phi}(N^{\mathcal{N}_2}_{p^*}), \hat{H}_{g,\phi}(N^{\mathcal{N}_1}_{p^*})\right]\right]...\right]\hat{\Pi}_{T_1}|\Psi\rangle+\epsilon_{\{\psi\}}\rule{101pt}{0pt}
\\
=\langle\Psi|\hat{\mathbb P}\left(\frac{i}{\hbar}\right)^{n-1}\left[\hat{\mathcal H}_{g,\phi}(\mathcal{N}_n),....\left[\hat{\mathcal H}_{g,\phi}(\mathcal{N}_3),\left[\hat{\mathcal H}_{g,\phi}(\mathcal{N}_2), \hat{\mathcal H}_{g,\phi}(\mathcal{N}_1)\right]\right]...\right]\hat{\Pi}_{T_1}|\Psi\rangle+O(\hbar)+\epsilon_{\{\psi\}}\rule{96pt}{0pt}
\\
=\Phi_n(\langle\hat{J}(T_1)\rangle,\langle\hat{h}(T_1)\rangle,\langle{\hat h}^{(\lambda)}(T_1)\rangle, \langle\hat{p}(T_1)\rangle,\langle\hat\alpha(T_1)\rangle, \mathcal{N}_i )+O(\hbar)+\epsilon_{\{\psi\}}\rule{201pt}{0pt}
\\
=\left\{{H}_{g,\phi}(\bar{N}_n),....\left\{{H}_{g,\phi}(\bar N_3),\left\{{H}_{g,\phi}(\bar{N}_2),{H}_{g,\phi}(\bar{N}_1)\right\}\right\}...\right\}\big|_{E_{I}^{a}(T_1), A_{b}^{J}(T_1),\phi( T_1), {P}(T_1), \bar{N}_i=\mathcal{N}_i}\rule{121pt}{0pt}
\\
+O(\hbar)+O(d^4)+\epsilon_{\{\psi\}}\rule{411pt}{0pt}
\end{split}
\end{equation}
where we replace $\hat{ H}_{g,\phi}(N^{\mathcal{N}_n}_{p^*})$ with $\hat{\mathcal H}_{g,\phi}(\mathcal{N}_n)$ using the spatial coordinate condition $(3.21)$.
The term $\epsilon_{\{\psi\}}$ represents the matter back reaction from the $\{\psi\}$ sector, and we will use $\epsilon_{\{\psi\}}$ to denote generic matter back reactions from the $\{\psi\}$ sector in the following. According to $(4.4)$, the emergent gravitational and $\phi$ fields satisfy:
\begin{equation}
\begin{split}
\\
H_{g,\phi}(\bar N)\big|_{E_{I}^{a}(T_1), A_{b}^{J}(T_1),\phi( T_1), {P}(T_1)}=0+O(\hbar)+O(d^4)+\epsilon_{\{\psi\}}\rule{55pt}{0pt}
\\\\
G_{g,\phi}(\bar{\Lambda})\big|_{E_{I}^{a}(T_1), A_{b}^{J}(T_1),\phi( T_1), {P}(T_1)}=0+O(\hbar)+O(d^4)+\epsilon_{\{\psi\}}\rule{55pt}{0pt}
\\\\
M_{g,\phi}(\bar V)\big|_{E_{I}^{a}(T_1), A_{b}^{J}(T_1),\phi( T_1), {P}(T_1)}=0+O(\hbar)+O(d^4)+\epsilon_{\{\psi\}}\rule{55pt}{0pt}
\\\\
\end{split}
\end{equation}
Therefore, the emergent gravitational and $\phi$ fields as a subsystem are on-shell up to corrections $O(\hbar)+O(d^4)+\epsilon_{\{\psi\}}$.

\subsection{ Homogeneous, Isotropic and Spatially Flat Dynamics}

 The coherent state $|\Psi\rangle$ gives the dynamics of $\{ E^{a}_{I}( X, T ), A^{J}_{b}( X, T), {\phi}(X, T), {P}(X,T)\}$, through the corresponding $\{\langle\hat{J}(e_{X, \Delta X}, T)_{I}\rangle,\langle\hat{h}(e_{X, \Delta X}, T)^{\bar{I}}_{\bar{J}}\rangle, \langle{\hat h}^{(\lambda)}(X, T) \rangle,\langle\hat{p}(X,T)\rangle\}$ at various $ T $. We now evaluate this clock time dynamics in a symmetrical setting.

Before the calculation, we need to set up a homogeneous, isotropic, and spatially flat initial condition at some late clock time $T_1$. Since the ${\phi}^0$ field serves as the clock, it is clear that $\langle{\hat\phi^0}(X, T_1)\rangle= T_1$, and $(3.13)$ implies that ${\phi}(X,T_1)$ is spatially constant.  We require the coherent state $|\Psi\rangle$ to give homogeneous $ E^{a}_{I}(X,T_1)$, $A^{J}_{b}(X,T_1)$ and $P(X, T_1)$, such that:
\begin{equation}
 E^{a}_{I}(X,T_1)= E_{1} \delta^a_I\rule{2pt}{0pt};\rule{4pt}{0pt}
 A^{J}_{b}(X,T_1)= A_{1}\delta^J_b\rule{2pt}{0pt};\rule{4pt}{0pt}P(X, T_1)=P_{1}\rule{2pt}{0pt};\rule{4pt}{0pt}{\phi}(X,T_1)=\phi_1
\end{equation}
where $\delta^a_I$ identifies $(x,y,z)$ with $(1,2,3)$. Note that $(4.6)$ satisfy $(4.5)$ as required, and it also implies that 
\begin{equation}
{\nu}_{\phi}(X, T_{1})={\nu_1}
\end{equation}

The dynamics of the emergent fields results from the symmetry of $\mathbb H$.
For any operator $\hat{O}(v)$ involving only gravitational and $\phi$ fields, $\frac{\partial}{\partial T}\langle \hat{O}( X, T)\rangle$ can be calculated using $\hat H(N_p) |\Psi\rangle=0$ to obtain \cite{lin1}:
\begin{equation}
\begin{split}
\frac{\partial}{\partial T}\bigg{|}_{T1}\langle\hat O(X, T)\rangle
= \langle\Psi| \hat{\mathbb P}\frac{i}{\hbar} \left[\hat O(X),\hat{H}\left(  N^{{\nu}^{-1}_1}_{p^{*}}\right)\right]\hat{\Pi}_{T_1}|\Psi\rangle+O(\hbar)\\
\end{split}
\end{equation} 
where the $O(\hbar)$ correction comes from the quantum fluctuation of the clock field\footnote{In general, there is also  a correction of $O(d)$ in $(4.8)$ due to the discretization error of inhomogeneous $\nu_\phi(X,T_1)$. In this paper, we have a homogeneous $\nu_\phi$ field at $T_1$, so the correction does not appear here.} \cite{lin1}.
To carry on, we first recall that our specific matter coordinates and frames satisfy $(3.30)$.
The correspondence $(3.34)$ implies that under a change of the spatial coordinates and frames, the emergent fields $\{ E^{a}_{I}( X, T ), A^{J}_{b}( X, T),\phi(X, T), {P}(X,T)\}$ transform classically by passive local $SU(2)$ and $\mathit{diff}_M$ transformations \cite{lin1}, up to errors of $O(\hbar)+O(d)$. Therefore, by applying the classical transformations to the result of our calculation, we can generalize it to arbitrary spatial coordinates and frames, up to errors of $O(\hbar)+O(d)$.

In the matter coordinates and frames satisfying $(3.30)$, the applications of $(4.8)$ lead to \cite{lin1}:
\begin{equation}
\begin{split}
\frac{d}{dT}\bigg|_{T_{1}} \langle \hat{J}(e_{X, \Delta X}, T)_{I} \rangle
=\frac{d}{dT}\bigg|_{T_{1}}\int_{\bar{S}_{X, \Delta X}}E^{a}_{I}(T)ds_{a}\rule{205pt}{0pt}\\
= \langle\Psi| \hat{\mathbb{P}}\frac{i}{\hbar} \left[ \hat{J}(e_{X, \Delta X})_{I}, \hat{H}_{g,{\phi}}\left(  N^{{\nu}^{-1}_{1}}_{p^{*}}\right)\right] \hat{\Pi}_{T_{1}}|\Psi\rangle +\epsilon_{\{\psi\}}+ O(\hbar)\rule{49pt}{0pt}\\
= \langle\Psi| \hat{\mathbb{P}}\frac{i}{\hbar} \left[  \hat{J}(e_{X, \Delta X})_{I} , \hat{\mathcal H}_{g,\phi}\left({\nu}^{-1}_{1}\right)\right] \hat{\Pi}_{T_{1}}|\Psi\rangle+\epsilon_{\{\psi\}}+ O(\hbar)\rule{64pt}{0pt}
\\\\
\frac{d}{dT}\bigg|_{T_{1}} \langle\hat{h}(e_{X, \Delta X}, T)^{\bar{I}}_{\bar{J}}\rangle
=\frac{d}{dT}\bigg|_{T_{1}} \mathcal{P}\exp[ \int_{\bar{e}_{X, \Delta X}} A^{J}_{b}(T)(\tau_{J})de^{b} ]^{\bar{I}}_{\bar{J}}\rule{152pt}{0pt}\\ 
= \langle\Psi| \hat{\mathbb{P}}\frac{i}{\hbar} \left[ \hat{h}(e_{X, \Delta X})^{\bar{I}}_{\bar{J}}, \hat{H}_{g,\phi}\left( N^{{\nu}^{-1}_{1}}_{p^{*}}\right)\right] \hat{\Pi}_{T_{1}}|\Psi\rangle+\epsilon_{\{\psi\}}+ O(\hbar)\rule{50pt}{0pt}\\
= \langle\Psi| \hat{\mathbb{P}}\frac{i}{\hbar} \left[  \hat{h}(e_{X, \Delta X})^{\bar{I}}_{\bar{J}},\hat{\mathcal H}_{g,\phi}\left( {\nu}^{-1}_{1}\right) \right] \hat{\Pi}_{T_{1}}|\Psi\rangle+\epsilon_{\{\psi\}}+ O(\hbar)\rule{66pt}{0pt}
\\\\
\frac{d}{dT}\bigg|_{T_{1}} \langle \hat{p}(X,T_1) \rangle
=\frac{d}{dT}\bigg|_{T_{1}}\int_{\bar{c}_{X}}{P}(X',T)dX'\rule{211pt}{0pt}\\
= \langle\Psi| \hat{\mathbb{P}}\frac{i}{\hbar} \left[ \hat{p}(X), \hat{H}_{g,\phi}\left( N^{{\nu}^{-1}_{1}}_{p^{*}}\right)\right] \hat{\Pi}_{T_{1}}|\Psi\rangle +\epsilon_{\{\psi\}}+ O(\hbar)\rule{82pt}{0pt}\\
= \langle\Psi| \hat{\mathbb{P}}\frac{i}{\hbar} \left[  \hat{p}(X) , \hat{\mathcal H}_{g,\phi}\left({\nu}^{-1}_{1}\right) \right] \hat{\Pi}_{T_{1}}|\Psi\rangle+\epsilon_{\{\psi\}}+ O(\hbar)\rule{97pt}{0pt}
\\\\
\end{split}
\end{equation}
where we again replace $\hat{ H}_{g,\phi}(N^{\mathcal{N}_n}_{p^*})$ with $\hat{\mathcal H}_{g,\phi}(\mathcal{N}_n)$ using the spatial coordinate condition $(3.21)$.
We then apply $(3.28)$,$(3.19)$ and $(3.31)$ to obtain:
\begin{equation}
\begin{split}
\frac{d}{dT}\bigg|_{T_{1}} \langle \hat{J}(e_{X, \Delta X}, T)_{I} \rangle
= \langle\Psi| \hat{\mathbb{P}}\frac{i}{\hbar} \left[  \hat{J}(e_{X, \Delta X})_{I} , \hat{\mathcal H}_{g,\phi^0}\left({{\nu}_{\phi^0}}^{-1}(T_1)\right)\right] \hat{\Pi}_{T_{1}}|\Psi\rangle+\epsilon_{\{\psi\}}+ O(\hbar)\rule{179pt}{0pt}\\
=\langle\Psi|(\Phi_{J}^{X, \Delta X})_{I}(\hat{J}(T_1),\hat{h}( T_1), {\hat h}^{(\lambda)}(T_1), \hat{p}(T_1))|\Psi\rangle\rule{250pt}{0pt}\\
+ \langle\Psi|(\Phi_{J,\alpha}^{X, \Delta X})_I(\hat{J}(T_1),\hat{h}(T_1), {\hat h}^{(\lambda)}(T_1), \hat{p}(T_1), \hat{\alpha}(T_1))|\Psi\rangle+\epsilon_{\{\psi\}}
+ O(\hbar)\rule{141pt}{0pt}
\\
=(\Phi_{J}^{X, \Delta X})_{I}(\langle\hat{J}(T_1)\rangle,\langle\hat{h}( T_1)\rangle,\langle {\hat h}^{(\lambda)}(T_1)\rangle, \langle\hat{p}(T_1)\rangle)\rule{248pt}{0pt}\\
+ (\Phi_{J,\alpha}^{X, \Delta X})_I(\langle\hat{J}(T_1)\rangle,\langle\hat{h}(T_1)\rangle, \langle{\hat h}^{(\lambda)}(T_1)\rangle,\langle \hat{p}(T_1)\rangle, \langle\hat{\alpha}(T_1)\rangle)+\epsilon_{\{\psi\}}
+ O(\hbar)\rule{130pt}{0pt}
\\\\
\frac{d}{dT}\bigg|_{T_{1}} \langle\hat{h}(e_{X, \Delta X}, T)^{\bar{I}}_{\bar{J}}\rangle
= \langle\Psi| \hat{\mathbb{P}}\frac{i}{\hbar} \left[  \hat{h}(e_{X, \Delta X})^{\bar{I}}_{\bar{J}},\hat{\mathcal H}_{g,\phi^0}\left( {{\nu}_{\phi^0}}^{-1}(T_1)\right) \right] \hat{\Pi}_{T_{1}}|\Psi\rangle+\epsilon_{\{\psi\}}+ O(\hbar)\rule{176pt}{0pt}\\
 =\langle\Psi|(\Phi_{h}^{X, \Delta X})^{\bar{K}}_{\bar{L}}(\hat{J}(T_1),\hat{h}(T_1), {\hat h}^{(\lambda)}(T_1), \hat{p}(T_1))|\Psi\rangle
+\epsilon_{\{\psi\}}+ O(\hbar)\rule{176pt}{0pt}\\
 =(\Phi_{h}^{X, \Delta X})^{\bar{K}}_{\bar{L}}(\langle\hat{J}(T_1)\rangle,\langle\hat{h}(T_1)\rangle, \langle{\hat h}^{(\lambda)}(T_1)\rangle,\langle \hat{p}(T_1)\rangle)
+\epsilon_{\{\psi\}}+ O(\hbar)\rule{174pt}{0pt}
\\\\
\frac{d}{dT}\bigg|_{T_{1}} \langle \hat{p}(X,T_1) \rangle
= \langle\Psi| \hat{\mathbb{P}}\frac{i}{\hbar} \left[  \hat{p}(X) , \hat{\mathcal H}_{g,\phi^0}\left({{\nu}_{\phi^0}}^{-1}(T_1)\right) \right] \hat{\Pi}_{T_{1}}|\Psi\rangle+\epsilon_{\{\psi\}}+ O(\hbar)\rule{199pt}{0pt}\\
 =\langle\Psi|\Phi_{p}^{X}(\hat{J}(T_1),\hat{h}(T_1), {\hat h}^{(\lambda)}(T_1), \hat{p}(T_1))|\Psi\rangle
+\epsilon_{\{\psi\}}+ O(\hbar)\rule{204pt}{0pt}\\
 =\Phi_{p}^{X}(\langle\hat{J}(T_1)\rangle,\langle\hat{h}(T_1)\rangle, \langle{\hat h}^{(\lambda)}(T_1)\rangle,\langle \hat{p}(T_1)\rangle)
+\epsilon_{\{\psi\}}+ O(\hbar)\rule{202pt}{0pt}
\\\\
\end{split}
\end{equation}

The first terms in $(4.10)$ are the $O(\hbar^0)$ contributions ignoring the back-reaction of $\{\psi\}$ sector. They are given by the functions $(\Phi_{J}^{X, \Delta X})_{I}$, $(\Phi_{J,\alpha}^{X, \Delta X})_{I}$, $(\Phi_{h}^{X, \Delta X})^{\bar{K}}_{\bar{L}}$ and $\Phi_{p}^{X}$, which are functions of the emergent fields through $(3.34)$. Recall that in our spatial coordinates $\Delta X_{m,i}\in\{(\pm d, 0,0), (0,\pm d,0), (0,0,\pm d)\}$ with $v^*_m\not \in \{v^*_{n_b}\}$, and define $\delta^{\Delta X}_{I}$ to have the only none-zero components:
\begin{equation}
\begin{split}
\delta^{(\pm d, 0,0)}_{1}= \pm 1\rule{2pt}{0pt}, \rule{4pt}{0pt}\delta^{( 0,\pm d,0)}_{2}= \pm 1\rule{2pt}{0pt}, \rule{4pt}{0pt} \delta^{( 0,0,\pm d)}_{3}= \pm 1
\nonumber
\end{split}
\end{equation}
For our symmetric case $(4.6)$ we have:
\begin{equation}
\begin{split}
(\Phi_{J}^{X, \Delta X})_{I}(\langle\hat{J}(T_1)\rangle,\langle\hat{h}( T_1)\rangle, \langle{\hat h}^{(\lambda)}(T_1)\rangle, \langle\hat{p}(T_1)\rangle)
=\delta^{\Delta X}_{I}\Phi_{J}( d^{2}E_{1},\exp({dA_{1}\tau_{i}}), \exp(i \lambda \phi_1), d^3 P_{1})\rule{25pt}{0pt}
\\\\
(\Phi_{J,\alpha}^{X, \Delta X})_{I}(\langle\hat{J}(T_1)\rangle,\langle\hat{h}(T_1)\rangle, \langle{\hat h}^{(\lambda)}(T_1)\rangle, \langle\hat{p}(T_1)\rangle, \langle\hat{\alpha}(T_1)\rangle)
=0\rule{185pt}{0pt}
\\\\
(\Phi_{h}^{X, \Delta X})^{\bar{K}}_{\bar{L}}(\langle\hat{J}(T_1)\rangle,\langle\hat{h}(T_1)\rangle,\langle{\hat h}^{(\lambda)}(T_1)\rangle, \langle\hat{p}(T_1)\rangle)=\delta^{\Delta X}_{I}(\tau^I)^{\bar{K}}_{\bar{L}}\Phi_{h}( d^{2}E_{1},\exp({dA_{1}\tau_{i}}), \exp(i \lambda\phi_1), d^3 P_{1})
\\\\
 \Phi_{p}^{X}(\langle\hat{J}(T_1)\rangle,\langle\hat{h}(T_1)\rangle, \langle{\hat h}^{(\lambda)}(T_1)\rangle, \langle\hat{p}(T_1)\rangle)= \Phi_{p}(d^{2}E_{1},\exp({dA_{1}\tau_{i}}), \exp(i \lambda \phi_1), d^3 P_{1})\rule{75pt}{0pt}
\\
\end{split}
\end{equation}
where the new set of functions $\{\Phi_{J}, \Phi_{h}, \Phi_{p}\}$ have no labels $X$, $\Delta X$ or $I$. The term $\Phi_{J,\alpha}^{X, \Delta X}$ vanishes because of the diagonal condition on $E^a_I$ in $(4.5)$. Refering to $(4.10)$ and $(4.9)$, we see  that $(4.11)$ implies the preservation $(4.6)$  along the clock time. Thus, at $O(\hbar^0)$ and ignoring the back reactions from $m'$, the dynamics of the gravitational and $\phi$ field is isotropic, homogeneous and spatially flat:
\begin{equation}
\begin{split}
 E^{a}_{I}(X,T)\equiv\delta^a_I {\boldsymbol E}(T) +\epsilon_{\{\psi\}}+ O(\hbar)\rule{1pt}{0pt};\rule{2pt}{0pt}
 A^{J}_{b}(X,T)\equiv\delta^J_b {\boldsymbol A}(T)+\epsilon_{\{\psi\}}+ O(\hbar)\\
P(X, T)\equiv{\boldsymbol P}(T)+\epsilon_{\{\psi\}}+ O(\hbar)\rule{1pt}{0pt};\rule{2pt}{0pt}\phi(X,T)\equiv\boldsymbol\phi (T)+\epsilon_{\{\psi\}}+ O(\hbar)\rule{20pt}{0pt}
\end{split}
\end{equation}
with the initial condition $(4.6)$: ${\boldsymbol E}(T_1)=E_1$, ${\boldsymbol A}(T_1)=A_1$, ${\boldsymbol P}(T_1)=P_1$ and $\boldsymbol\phi(T_1)=\phi_1$.

Recall from $(4.5)$, the emergent fields satisfy the Gauss, momentum and Hamiltonian constraints up to the corrections. In our specific case, the Gauss and momentum constraints are trivially satisfied while the Hamiltonian constraint gives nontrivial implications for the dynamics. Inserting $(4.12)$ into $(4.4)$ with $n=1$, we have:
\begin{equation}
\begin{split}
\Phi_1( d^2{\boldsymbol E}(T),\exp(d {\boldsymbol A}(T)\tau_{J}),\exp(i \lambda \boldsymbol\phi(T)), d^3{\boldsymbol P}(T), \mathcal{N})+\epsilon_{\{\psi\}}+O(\hbar)\\=H_{g,\phi}(\bar N)\big|_{ \delta^a_I {\boldsymbol E}(T) , \delta^J_b {\boldsymbol A}(T),\boldsymbol\phi( T), {\boldsymbol P}(T), \bar{N}=\mathcal{N}}+O(d^4)+\epsilon_{\{\psi\}}+O(\hbar)\rule{34pt}{0pt}\\=0\rule{311pt}{0pt}
\end{split}
\end{equation}
Here we see that the $O(\hbar^0)$ effective Hamiltonian constraint for gravitational and $\phi$ field is given by $\Phi_1$, which contains corrections of $O(d^4)$ to the classical $H_{g,\phi}$.

To describe this symmetric sector in a symmetrically reduced form, we introduce the symmetrically reduced variables -- $\{\boldsymbol p,\boldsymbol h_{i}\equiv \exp( \boldsymbol c\tau_i)\}$ $(i=1,2,3)$ in place of $\{d^2 \boldsymbol E,\exp(d\boldsymbol A\tau_{I})\}$, and $\{\boldsymbol p_{\phi} , \boldsymbol h_{\phi}\equiv \exp(i \lambda_0 \boldsymbol\phi)\}$ in place of $\{d^3 \boldsymbol P, \exp(i \lambda_0 \boldsymbol \phi)\}$. Their non-zero Poisson brackets are defined by:
\begin{equation}
\{ \boldsymbol p, \boldsymbol c\} \equiv \frac{1}{3}\kappa\gamma \rule{5pt}{0pt}; \rule{20pt}{0pt} \{\boldsymbol p_{\phi}, \boldsymbol\phi\}= 1
\end{equation}
where the factor $\frac{1}{3}$ accounts for the degeneracy on the three independent spatial directions. We now define $H_{lqc}(\boldsymbol N)$ $(\boldsymbol N\in \mathbb R)$ to be a function of $\{\boldsymbol p, \boldsymbol h_{i},\boldsymbol p_{\phi},\boldsymbol h_{\phi}\}$ that satisfies:
\begin{equation}
H_{lqc}(\boldsymbol N)\big{|}_{\boldsymbol p=d^{2}\boldsymbol E,\boldsymbol h_{i}=\exp(d \boldsymbol A\tau_{i}),\boldsymbol p_{\phi}=d^3 \boldsymbol P,\boldsymbol h_{\phi}=\exp(i \lambda_0 \boldsymbol\phi) }=\Phi_1( d^2\boldsymbol E,\exp(d \boldsymbol A\tau_{J}),\exp(i \lambda_0 \boldsymbol\phi), d^3\boldsymbol P, \boldsymbol N)\rule{0pt}{0pt}
\end{equation}
 Its explicit form is given by:
\begin{equation}
\begin{split}
H_{lqc}(\boldsymbol N)
\equiv{ H}^E_{lqc}(\boldsymbol N)\rule{400pt}{0pt}\\
 - \boldsymbol N\cdot 2(1+\gamma^2)\frac{2}{\kappa^4\gamma^7} \sum_{i,j,k=1}^{3} \epsilon^{ijk}    
  tr \left[{\boldsymbol h^{-1}_{i}}\{{\boldsymbol h}_i,\{H^{E}_g(1),p^{\frac{3}{2}}\} \}{\boldsymbol h^{-1}_{j}}\{\hat{\boldsymbol h}_j,\{H^{E}_g(1),p^{\frac{3}{2}}\} \}  {\boldsymbol h}_k\{{ \boldsymbol h^{-1}_{k}}, \boldsymbol p^{\frac{3}{2}}\}\right]\rule{10pt}{0pt}\\
+N\cdot2\kappa \boldsymbol p^{-3/2}\boldsymbol  p_{\phi}^2 \rule{365pt}{0pt}
\\\\
\end{split}
\end{equation}
where
\begin{equation}
\begin{split}
H^{E}_{lqc}(\boldsymbol N)\equiv\boldsymbol N  \frac{2}{2\kappa^2\gamma}\sum_{i,j,k=1}^{3}\epsilon^{ijk}  tr \left[({\boldsymbol h}_{ij}-{\boldsymbol h}_{ji})\hat{\boldsymbol h^{-1}_{k}}\{\hat{\boldsymbol h}_k, \boldsymbol p^{\frac{3}{2}} \}\right]
\nonumber
\end{split}
\end{equation}
Next, we define:
\begin{equation}
\begin{split}
\boldsymbol\phi^0=\frac{1}{2i\lambda_0}[\boldsymbol h_{\phi}-\boldsymbol h^*_{\phi}]\rule{2pt}{0pt};\rule{4pt}{0pt}\boldsymbol v_{\phi^0}\equiv\{\boldsymbol\phi^0,H_{lqc}(1)\}
\nonumber
\end{split}
\end{equation}
and find that the equations of motion $(4.12)$ can be expressed as
\begin{equation}
\begin{split}
\frac{d}{dT}\bigg|_{T_1}  d^2 \boldsymbol E(T)
= \left\{{\boldsymbol p},H_{lqc}(\boldsymbol N)\right\}\bigg{|}_{\boldsymbol N=\boldsymbol{v}_{\phi^0}^{-1};\rule{4pt}{0pt}\boldsymbol p=d^{2}\boldsymbol E(T_1),\boldsymbol h_{i}=\exp({d\boldsymbol A(T_1)\tau_{i}}),\boldsymbol p_{\phi}=d^3 \boldsymbol P(T_1), \boldsymbol h_{\phi}(T_1)=\exp(i \lambda_0 \boldsymbol\phi(T_1))}\\
+\epsilon_{\{\psi\}}+ O(\hbar)\rule{293pt}{0pt}
\\\\
\frac{d}{dT}\bigg|_{T_1}  \exp({d\boldsymbol A(T)\tau_{i}})
= \left\{{\boldsymbol h}_i,H_{lqc}(\boldsymbol N)\right\}\bigg{|}_{\boldsymbol N=\boldsymbol{v}_{\phi^0}^{-1};\rule{4pt}{0pt}\boldsymbol p=d^{2}\boldsymbol E(T_1),\boldsymbol h_{i}=\exp({d\boldsymbol A(T_1)\tau_{i}}),\boldsymbol p_{\phi}=d^3 \boldsymbol P(T_1), \boldsymbol h_{\phi}(T_1)=\exp(i \lambda_0 \boldsymbol\phi(T_1))}
\\+\epsilon_{\{\psi\}}+ O(\hbar)\rule{293pt}{0pt}
\\\\
\frac{d}{dT}\bigg|_{T_1} d^3 \boldsymbol P(T)
=\left\{\boldsymbol p_{\phi},H_{lqc}(\boldsymbol N)\right\}\bigg{|}_{\boldsymbol N=\boldsymbol{v}_{\phi^0}^{-1};\rule{4pt}{0pt}\boldsymbol p=d^{2}\boldsymbol E(T_1),\boldsymbol h_{i}=\exp({d\boldsymbol A(T_1)\tau_{i}}),\boldsymbol p_{\phi}=d^3 \boldsymbol P(T_1), \boldsymbol h_{\phi}(T_1)=\exp(i \lambda_0 \boldsymbol\phi(T_1))}
\\
+\epsilon_{\{\psi\}}+ O(\hbar)\rule{293pt}{0pt}
\\\\
\end{split}
\end{equation}
Equations $(4.17)$ state that the $ O(\hbar^0)$ evolution of the emergent gravitational and $\phi$ fields in the state $|\Psi\rangle$ is governed by the effective Hamiltonian constraint $H_{lqc}(N)$, ignoring matter back reactions from $\{\psi\}$ sector. To this approximation, the symmetrically reduced model captures the clock time dynamics of the emergent fields through $(\boldsymbol p(\boldsymbol\phi^0),\boldsymbol c(\boldsymbol\phi^0),\boldsymbol p_{\phi}(\boldsymbol\phi^0))=(d^2\boldsymbol E(T), d\boldsymbol A(T),d^3\boldsymbol P(T))$, and it is straightforward to check that we have a conserved observable $d^3\boldsymbol P(T)= const=d^3 P_1$.

Next, we investigate the evolution of the spatial scale.
 Recall from $(2.16)$ that the volume operator with a dynamical region $R$ is obtained by summing over the operators $\sqrt{|\hat q_{v_n}|}$  with $ v_n\in R$. Using $(3.15)$ and $(3.17)$, we can localize $\sqrt{|\hat q_{v_n}|}$ as $\sqrt{|\hat q(X_n, T)|}$. Therefore, the spatial volume observable corresponding to the coordinate region $\bar\Omega\subset \bar I^3$ at a clock time $T$ is given by:
\begin{equation}
\hat V( \bar\Omega , T)\equiv\sum_{X_m\in \bar\Omega}\sqrt{|\hat q(X_m, T)|} 
\nonumber
\end{equation}
To investigate the evolving spatial scale in the cosmology, we now keep track of the volume of the spatial region coordinatized by $\bar c_{X_m }$ at various clock times:
\begin{equation}
\begin{split}
\frac{\partial}{\partial T}\bigg|_{T_{1}} \langle \hat{V}(\bar c_{X_m },T) \rangle
=\frac{\partial}{\partial T}\bigg|_{T_{1}}   \int_{ \bar c_{X_m}}\sqrt{\frac{1}{3!}\epsilon^{IJK}\epsilon_{abc}{E}^{a}_{I}{E}^{b}_{J}{E}^{c}_{K}(X',T)} d^3X'+ O(\hbar)\rule{200pt}{0pt}
\\
=\frac{d}{dT}\bigg|_{T_{1}}  d^3 \boldsymbol E^{3/2}(T)+ O(\hbar)\rule{341pt}{0pt}
\\
= \left\{\boldsymbol p^{3/2},H_{lqc}(\boldsymbol N)\right\}\bigg{|}_{\boldsymbol N=\boldsymbol{v}_{\phi^0}^{-1};\rule{4pt}{0pt}\boldsymbol p=d^{2}\boldsymbol E(T_1),\boldsymbol h_{i}=\exp({d\boldsymbol A(T_1)\tau_{i}}),\boldsymbol p_{\phi}=d^3 \boldsymbol P(T_1), \boldsymbol h_{\phi}(T_1)=\exp(i \lambda_0 \boldsymbol\phi(T_1))}\rule{81pt}{0pt}\\
+\epsilon_{\{\psi\}}+ O(\hbar)\rule{390pt}{0pt}
\\
=-\frac{3}{2\gamma \boldsymbol v_{\phi^0}(T_1)}d^{2}\boldsymbol E(T_1)\left[2\sin(d\boldsymbol A(T_1)\cos(d\boldsymbol A(T_1)\left[1-2(1+\gamma^2)\sin^2(d\boldsymbol A(T_1)\right]\right]\rule{99pt}{0pt}\\
+\epsilon_{\{\psi\}}+ O(\hbar)\rule{390pt}{0pt}
\\
\end{split}
\end{equation}
Also, the constraint equation $(4.13)$ leads to:
\begin{equation}
\begin{split}
\sin^2(d\boldsymbol A(T_1))\left[1-(1+\gamma^2)\sin^2(d\boldsymbol A(T_1))\right]+\epsilon_{\{\psi\}}+ O(\hbar)
=\frac{4\kappa^2\gamma^2(d^3\boldsymbol P(T_1))^2}{6d^4\boldsymbol E^{2}(T_1)}
\equiv\left(\frac{\boldsymbol \rho_{\phi}(T_1)}{\boldsymbol \rho_c}\right)^{\frac{2}{3}}
\rule{0pt}{0pt}
\end{split}
\end{equation}
Since the energy density of the $\phi$ field is given by $\boldsymbol \rho_{\phi}(T)=\frac{ 2\kappa\boldsymbol  P^2(T)}{\boldsymbol E^{3}(T)}$, $\boldsymbol \rho_c$ is equal to the conserved quantity $(\frac{2}{27}\kappa^4\gamma^6)^{-1/2}(d^3\boldsymbol  P(T))^{-1} =(\frac{2}{27}\kappa^4\gamma^6)^{-1/2}(d^3 P_1 )^{-1} $.

To obtain our modified first Friedmann equation, we solve the constraint equation $(4.19)$ and find:
\begin{equation}
\begin{split}
\sin^2(d\boldsymbol A(T_1))
=\frac{1-\sqrt{1-\boldsymbol \chi(T_1)}}{2(1+\gamma^2)}
+\epsilon_{\{\psi\}}+ O(\hbar)\rule{2pt}{0pt};\rule{4pt}{0pt}
\boldsymbol \chi(T_1)\equiv 4(1+\gamma^2){\left(\frac{\boldsymbol \rho_{\phi}(T_1)}{\boldsymbol \rho_c}\right)}^{\frac{2}{3}}\rule{0pt}{0pt}\\
\end{split}
\end{equation}
Substituting $(4.20)$ into $(4.18)$ we obtain the first modified Friedmann equation for the Hubble constant $\boldsymbol H(T)$:
\begin{equation}
\begin{split}
\boldsymbol H^2(T_1)\rule{382pt}{0pt}\\
=\left[\frac{\boldsymbol v_{\phi^0}(T_1)\frac{d}{dT}\big|_{T_{1}} {\boldsymbol E}^{\frac{3}{2}}(T) }{3{\boldsymbol E}^{\frac{3}{2}}(T_1) }\right]^2\rule{306pt}{0pt}
\\
=\frac{d^4}{4\gamma^2(1+\gamma^2)^2\boldsymbol E(T_1)}( 1-\boldsymbol \chi(T_1))\left( 1-\sqrt{1-\boldsymbol \chi(T_1)}\right)\left( 1+2\gamma^2+\sqrt{1-\boldsymbol\chi(T_1)}\right)+\epsilon_{\{\psi\}}+ O(\hbar)\rule{0pt}{0pt}\\
\end{split}
\end{equation}
When the universe is in the classical region, we have $\boldsymbol\rho_c \gg \boldsymbol\rho_{\phi}(T_1)$ and $\boldsymbol\chi(T_1)\ll 1$, and $(4.21)$ approaches the classical first Friedmann equation:
\begin{equation}
\begin{split}
\boldsymbol H^2(T_1)
=\frac{\kappa}{3}\boldsymbol\rho_{\phi}(T_1)+\epsilon_{\{\psi\}}+ O(\hbar)\\
\end{split}
\end{equation}
When $\boldsymbol\rho_{\phi}$ becomes comparable to $\boldsymbol\rho_c$, our model gives significant modifications to the classical FRW cosmology.  It is clear from $(4.21)$ that $\boldsymbol H(T_1)$ vanishes when $\boldsymbol\rho_{\phi}(T_1)=\boldsymbol\rho_c$ and that the initial singularity is replaced by a  bouncing behavior, up to the corrections of $\epsilon_{\{\psi\}}+ O(\hbar)$.

\section{Comparisons with Loop Quantum Cosmology at $O(\hbar^0)$}

 We have now identified a symmetric sector of the model described in the first half of the paper. This sector is represented by a state $|\Psi\rangle$ which gives an approximately homogeneous, isotropic and spatially flat evolution of the emergent gravitational and $\phi$ fields.  Further, the evolution is effectively governed by the symmetrically reduced classical Hamiltonian constraint $H_{lqg}(\boldsymbol N)$. Also, the evolution of the fields agrees with FRW cosmology in large scales, while it deviates from FRW cosmology when the $\phi$ field energy density is close to the critical value $\boldsymbol\rho_c$. Finally, the deviation results in the resolution of the initial singularity. In the following, we will compare $H_{lqg}(\boldsymbol N)$ with the effective Hamiltonian constraints of $O(\hbar^0)$ in the existing models of loop quantum cosmology.

The homogeneous, isotropic and spatially flat sector of classical general relativity can be described by a symmetrically reduced theory, namely the flat FRW cosmology.  In our setting, the flat FRW cosmology describes an arbitrary representative cell $\mathcal C$ of the comoving space with four (off-shell) phase space variables $\{\boldsymbol p,\boldsymbol c, \boldsymbol\phi,\boldsymbol p_{\phi}\}$. Since the space is flat, we can choose an arbitrary Euclidean spatial coordinate system for $\mathcal C$, in which $\mathcal C$ has a coordinate volume $\mathcal V$. In this spatial coordinate system, the reduced phase space variables are related to the Ashtekar's variables by \cite{lqcint2,lqcint3,lqcint4}:
\begin{equation}
\mathcal V^{1/3}A^{i}_{a}(\text x)=\boldsymbol c\cdot\omega^{i}_{a}\rule{3pt}{0pt};\rule{10pt}{0pt} \mathcal V^{2/3} E_{j}^{b}(\text x)= \boldsymbol p\cdot e^{b}_{j}\rule{3pt}{0pt};\rule{10pt}{0pt}\mathcal{V} P(\text x)=\boldsymbol p_{\phi};\rule{10pt}{0pt}\phi(\text x)=\boldsymbol\phi
\end{equation}
The triads $e^{b}_{j}$ and cotriads $\omega^{i}_{a}$ are non-dynamical, and they will be gauge-fixed to identity matrices from now on. The two conjugate pairs satisfy the Poisson brackets:
\begin{equation}
 \{\boldsymbol c,\boldsymbol p\}=\frac{1}{3}\kappa\gamma ;\rule{10pt}{0pt}
\{\boldsymbol\phi,\boldsymbol p_{\phi}\}=1
\nonumber
\end{equation}
For our case, the Hamiltonian constraint $H_{FRW}(\boldsymbol N)$ for flat FRW cosmology is given by \cite{lqcint2,lqcint3,lqcint4}:
\begin{equation}
H_{FRW}(\boldsymbol N)= -\boldsymbol N\cdot\frac{3}{\kappa\gamma^2}\boldsymbol c^2 \boldsymbol p^{1/2}+ N\cdot2\kappa \boldsymbol p^{-3/2}\boldsymbol p_{\phi}^2 
\end{equation}
where $\boldsymbol N\in \mathbb R$ is the symmetrically reduced lapse function.

In the spirit of loop quantum gravity, loop quantum cosmology uses $\{ \boldsymbol p, \boldsymbol h^{\mu}_i\equiv \exp(\mu \boldsymbol c\tau_i)\}$ as elementary gravitational variables instead of $\{\boldsymbol p,\boldsymbol c\}$. The holonomy $ \boldsymbol h^{\mu}_i$ is given by the parallel transportation by the connection field $A^{i}_{a}$ along a certain path in the direction of $e^a_i$. In the Euclidean spatial coordinate system used in $(5.1)$, the coordinate length of the path is set to be $\mu\mathcal V^{1/3}$. Each model in homogeneous, isotropic and spatially flat loop quantum cosmology has an effective Hamiltonian constraint of $O(\hbar^0)$ that approximates $H_{FRW}(\boldsymbol N)$ in terms of the loop variables $\{ \boldsymbol p,\boldsymbol h^{\mu}_i\}$. Clearly, there are ambiguities in such approximations, and the different approximation schemes based on different physical considerations result to the different  models in loop quantum cosmology \cite{lqc0,earlylqc,lqcint1,lqcint2,lqcint3,lqcint4,altlqc}.

The first main ambiguity lies in the ordering of the two required procedures in obtaining the effective Hamiltonian constraints of  loop quantum cosmology from the full form $H_{g,\phi}(\bar N)$: symmetry reduction of the phase space and the introduction of the loop variables. When the symmetry reduction is applied first \cite{lqcint2,lqcint3,lqcint4}, $H_{g,\phi}(\bar N)$ is first simplified to the gravitational term in $H_{FRW}(\boldsymbol N)$, before the replacement of $\{\boldsymbol p,\boldsymbol c\}$ by $\{ \boldsymbol p, \boldsymbol h^{\mu}_i\}$. When the loop variables are introduced first \cite{altlqc}, $H_{g,\phi}(\bar N)$ is first approximated by the local holonomy and flux variables, before imposing the symmetry on the local loop variables and reducing them into the set $\{ \boldsymbol p, \boldsymbol h^{\mu}_i\}$. The first procedure leads to a simpler effective constraint, while the second leads to an effective constraint closer to the full form. We will denote the two schemes as $I$ and $II$.

The second main ambiguity lies in the choice of $\mathcal C$ and $\mu$. In FRW cosmology, the choice of $\mathcal C$ does not effect the physical predictions for $\{A^{i}_{a}, E_{j}^{b}, \phi,P\}$. This is because of the consistent scaling of $\{\boldsymbol p,\boldsymbol c,\boldsymbol \phi,\boldsymbol p_{\phi}\}$ and $H_{FRW}(\boldsymbol N)$ when $\mathcal C$ is changed. Given the scaling of $\{\boldsymbol p,\boldsymbol c,\boldsymbol \phi,\boldsymbol p_{\phi}\}$ with $\mathcal V$ according to $(5.1)$, $H_{FRW}(\boldsymbol N)$ scales linearly with $\mathcal V$ as supposed. In loop quantum cosmology, this independence from the choice of $\mathcal C$ may be disrupted due to the replacement of of $\boldsymbol c$ by $\boldsymbol h^{\mu}_i$. Particularly, the effective hamiltonian constraints of loop quantum cosmology can scale nonlinearly with $\mathcal V$ when $\mathcal C$ is changed. Therefore, the choice of $\mathcal C$ does affect the physics in some models loop quantum cosmology.

Given a choice of $\mathcal C$, the value of $\mu$ must also be specified. Clearly, the value of $\mu \mathcal{V}^{1/3}$ must be small enough that the effective Hamiltonian constraint can give a good approximation of $H_{FRW}$. In the early stage \cite{earlylqc} of loop quantum cosmology $\mu$ is set to be a small fixed constant $\mu=\mu_0$. However, two main objections were raised against this choice \cite{lqc0}. First, with $\mu=\mu_0$, the $O(\hbar^0)$ effective Hamiltonian constraint scales nonlinearly with $\mathcal{V}$ under a change of $\mathcal C$. This is a problem for loop quantum cosmology since the physics of $\{A^{i}_{a}, E_{j}^{b}, \phi,P\}$ depends on the choice of $\mathcal{C}$, while there is no natural way of making such a choice. Second, with $\mu=\mu_0$, the critical $\phi$ field energy density $\rho_c$ at the big bounce is inversely proportional to the conserved quantity $P$. Thus there is a danger of predicting the big bounce at a low matter density if $P$ is high enough. This is a problem since a high $P$ value is preferred for semi-classical limits. Because of these objections, a physically motivated choice with $\mu=\bar{\mu}\equiv D \boldsymbol p^{-1/2}$ has been adopted \cite{lqc0,lqcint1,lqcint2}, where $D$ is a real fixed parameter. The appearance of $ \boldsymbol p$ in $\mu$ restores the independence of $\mathcal C$ for loop quantum cosmology with $\mu=\bar\mu$.  Moreover, it also leads to a fixed critical density set at the Plank scale. We will denote the old and new choices in $\mu$ and $\mathcal C$ as $(\mathcal C, \mu_0)$ and $(\mathcal C, \bar\mu)$ schemes.

The combinations of these schemes give four distinct versions of the $O(\hbar^0)$ effective hamiltonian constraint $H_{{}_{LQC}}$. We will use the superscripts $(\mathcal C, \mu_0, I)$, $(\mathcal C, \mu_0, II)$, $(\mathcal C, \bar\mu, I)$ and $(\mathcal C, \bar\mu, II)$ to indicate the schemes applied. First, we define the $H^{E}_{{}_{LQC}}$  with a general choice of $\mu$ as
\begin{equation}
H^{E,\mu}_{{}_{LQC}}(\boldsymbol N)\equiv \boldsymbol N  \frac{2}{2(\mu)^{3}\kappa^2\gamma}\sum_{i,j,k=1}^{3}\epsilon^{ijk}  tr \left[(\boldsymbol{h}^\mu_{ij}-\boldsymbol{h}^\mu_{ji}){\boldsymbol h^{-1\mu}_{k}}\{\boldsymbol{h}^\mu_k,\boldsymbol p^{\frac{3}{2}} \}\right]\rule{190pt}{0pt}
\end{equation}
 The two effective Hamiltonian constraints using $\mu=\mu_0$ are:
\begin{equation}
 \begin{split}
   {H}^{(\mathcal C, \mu_0, I)}_{{}_{LQC}}(\boldsymbol N)
=  -\gamma^{-2}H^{E,\mu_0}_{{}_{LQC}}(\boldsymbol N)+ \boldsymbol N 2\kappa \boldsymbol p^{-3/2}\boldsymbol p_{\phi}^2
\rule{280pt}{0pt}
\\\\
H^{(\mathcal C, \mu_0, II)}_{{}_{LQC}}(\boldsymbol N)
\equiv H^{E,\mu_0}_{{}_{LQC}}(\boldsymbol N)\rule{380pt}{0pt}\\
 - \frac{4\boldsymbol N(1+\gamma^2)}{\kappa^4\gamma^7} \sum_{i,j,k=1}^{3} \epsilon^{ijk}    
  tr \left[{\boldsymbol h^{-1\mu_0}_{i}}\{\boldsymbol{h}^{\mu_0}_i,\{H^{E,\mu_0}_{{}_{LQC}}(1),\boldsymbol p^{\frac{3}{2}}\} \}{\boldsymbol h^{-1\mu_0}_{j}}\{\boldsymbol{h}^{\mu_0}_j,\{H^{E,\mu_0}_{{}_{LQC}}(1),\boldsymbol p^{\frac{3}{2}}\} \}  \boldsymbol{h}^{\mu_0}_k\{{\boldsymbol h^{-1\mu_0}_{k}}, \boldsymbol p^{\frac{3}{2}}\}\right]\rule{30pt}{0pt}\\
+ \boldsymbol N 2\kappa\boldsymbol p^{-3/2}\boldsymbol p_{\phi}^2 \rule{435pt}{0pt}
\\\\
\end{split}
\end{equation}
The other two using $\mu=\bar\mu\equiv D \boldsymbol p^{-1/2}$:
\begin{equation}
 \begin{split}
{H}^{(\mathcal C, \bar\mu, I)}_{{}_{LQC}}(\boldsymbol N)
=  -\gamma^{-2}H^{E, \bar\mu}_{{}_{LQC}}(\boldsymbol N)+ \boldsymbol N 2\kappa \boldsymbol p^{-3/2}\boldsymbol p_{\phi}^2
\rule{330pt}{0pt}
\\\\
H^{(\mathcal C, \bar\mu, II)}_{{}_{LQC}}(\boldsymbol N)
\equiv H^{E,\bar\mu}_{{}_{LQC}}(\boldsymbol N)\rule{430pt}{0pt}\\
 - \frac{4}{9}\cdot\frac{4\boldsymbol N(1+\gamma^2)}{\kappa^4\gamma^7} \sum_{i,j,k=1}^{3} \epsilon^{ijk}    
  tr \left[{\boldsymbol h^{-1\bar\mu}_{i}}\{\boldsymbol{h}^{\bar\mu}_i,\{H^{E,\bar\mu}_{{}_{LQC}}(1),\boldsymbol p^{\frac{3}{2}}\} \}{\boldsymbol h^{-1\bar\mu}_{j}}\{\boldsymbol{h}^{\bar\mu}_j,\{H^{E,\bar\mu}_{{}_{LQC}}(1),\boldsymbol p^{\frac{3}{2}}\} \}  \boldsymbol{h}^{\bar\mu}_k\{{\boldsymbol h^{-1\bar\mu}_{k}}, \boldsymbol p^{\frac{3}{2}}\}\right]\rule{90pt}{0pt}\\
+ \boldsymbol N 2\kappa\boldsymbol p^{-3/2}\boldsymbol p_{\phi}^2 \rule{483pt}{0pt}
\\\\
\end{split}
\end{equation}
One can easily check that 
\begin{equation}
\begin{split}
\\
H_{FRW}(\boldsymbol N)
=\lim_{\mu_0\mathcal{V}^{1/3}\to 0}{H}^{(\mathcal C, \mu_0, I)}_{{}_{LQC}}(\boldsymbol N)=\lim_{\mu_0\mathcal{V}^{1/3}\to 0}{H}^{(\mathcal C, \mu_0, II)}_{{}_{LQC}}(\boldsymbol N)=\lim_{\bar\mu\mathcal{V}^{1/3}\to 0}{H}^{(\mathcal C, \bar\mu, I)}_{{}_{LQC}}(\boldsymbol N)\\=\lim_{\bar\mu\mathcal{V}^{1/3}\to 0}{H}^{(\mathcal C, \bar\mu, II)}_{{}_{LQC}}(\boldsymbol N) \rule{233pt}{0pt}
\end{split}
\end{equation}
Note that $(5.3)$ and $(5.4)$ have almost the same form except for the extra $4/9$ factor in $H^{(\mathcal C, \bar\mu, II)}_{{}_{LQC}}$, and their qualitative difference lies in that $\bar\mu$ depends on $\boldsymbol p$.

The simplest $(\mathcal C, \mu_0, I)$ scheme was studied in the early stages \cite{earlylqc} of the program. The improved scheme $(\mathcal C, \bar\mu, I)$ has been extensively studied both analytically and numerically \cite{lqcint1,lqcint2,lqcint3,lqcint4}. The $(\mathcal C, \bar\mu, II)$ scheme that is closer to the full form of loop quantum gravity has recently been investigated \cite{altlqc}. Lastly, the effective model we $\mathit{derived}$ from the $O(\hbar^0)$ contributions of the semi-classical limit of the full theory corresponds to the $(\mathcal C, \mu_0, II)$ scheme. Denoting ${ c}_{X_m}$ as the region of the emergent space coordinatized by $\bar{ c}_{X_m}$, we compare $(4.15)$ and $(4.16)$ to $(5.4)$ and find:
\begin{equation}
\begin{split}
H^{(\mathcal C, \mu_0, II)}_{{}_{LQC}}(\boldsymbol N)\big|_{\mathcal{ C}={ c}_{X_m}, \mu_0=1}=H_{lqc}(\boldsymbol N)\\
\end{split}
\end{equation}

While the different schemes are physically distinct, they share important qualitative features, which represent the most robust part of loop quantum cosmology. The first three schemes consistently resolve initial singularity by a bouncing of the scale factor, as shown in \cite{lqcint1,lqcint2,lqcint3,lqcint4,altlqc} to various approximations. In the approximation of $O(\hbar^0)$, we have seen that the semi-classical limit given by the state $|\Psi\rangle$ in our mode matches with the $(\mathcal C, \mu_0, II)$ scheme, which also predicts the same bouncing behavior. From $(4.20)$ and $(4.21)$ it is clear that when the density $\boldsymbol\rho_{\phi}$ is close to the critical density $\boldsymbol\rho_c$, the contributions from $\sin(d\boldsymbol A(T))$ that are nonlinear in $d\boldsymbol A(T)$ become important, and drive the evolution away from the FRW cosmology. At $O(\hbar^0)$ level, the big bounces in the four schemes are caused by these non-linear holonomy corrections. From $(5.6)$ we see that the holonomy corrections appears only with non-zero values of $\mu \mathcal{V}^{1/3}$, which via $(5.7)$ corresponds to the non-zero value for $d$ in our model. Since the non-vanishing $d$ in our model reflects the intrinsic discreteness of space, our model supports the idea that the quantum geometry leads to the big bounce.

It should be emphasized that we obtained the $(\mathcal C, \mu_0, II)$ scheme from the spatial quantum geometry of loop quantum gravity. Through our model, the first-principle derivation not only leads to the specific scheme, but also determines the values of $\mathcal C$ and $\mu_0$ with fundamental meanings.
To see this, let us look at the objections to the $\mu_0$ schemes in light of our model. Recall that the first objection involes the sensitivity of the $\mu_0$ schemes to the arbitrary representative cell $\mathcal C$ in the space. From $(5.7)$, it is clear that such sensitivity corresponds to our model's sensitivity to $ c_{X_m}$. However, the objection does not apply to our model, where $ c_{X_m}$ is not arbitrary, but is an elementary cell in the space that is dual to a single physical node. The elementary cell represents one of the smallest regions in the emergent space with nonzero volume. Conversely, our model states that the cell $\mathcal C$ in the $\mu_0$ scheme models corresponds to one of the elementary cells of the emergent space in the full theory, and that $\mu_0=1$ so the holonomy variables simply run over the sides of this elementary cell.

The second objection involves the possibility of the big bounce happening at a low matter density, or at a large volume of space. In the context of our model, one can explicitly check whether this really occurs for the state $|\Psi\rangle$ describing our universe. Through the model, we now make a trial estimation of the scale of the big bounce using known cosmological factors. Recall from $(4.19)$, $(4.20)$ and $(4.21)$ that the $\phi$ field energy density is $\boldsymbol\rho_{\phi}(T)=\frac{ 2\kappa P^2_1}{\boldsymbol E^{3}(T)}$ and the critical energy density is $\boldsymbol\rho_c=(\frac{2}{27}\kappa^4\gamma^6)^{-1/2}(d^3P_1 )^{-1}$. At the moment $T_c$ of the big bounce we have:
\begin{equation}
\boldsymbol\rho_{\phi}(T_c)=\frac{ 2\kappa (d^3P_1)^2}{(d^2\boldsymbol E(T_c))^{3}}=(\frac{2}{27}\kappa^4\gamma^6)^{-1/2}(d^3P_1 )^{-1}=\boldsymbol\rho_c 
\end{equation}
On the other hand, at current cosmological time $T_1$ we have:
\begin{equation}
\boldsymbol\rho_{\phi}(T_1)=\frac{ 2\kappa (d^3P_1)^2}{(d^2\boldsymbol E(T_1))^{3}}
\end{equation}
In our setting with negligible energy density of the $\{\psi\}$ sector, $\boldsymbol\rho_{\phi}(T_1)$ is approximately equal to the dark energy density observed today, which is $\boldsymbol\rho_{\phi}(T_1)\approx 10^{-9} J/m^3$. Here we will use the value of Immirzi parameter $\gamma\approx \ln(2)/\pi$, given by black hole entropy consideration in loop quantum gravity \cite{gamma}. Also, since no dispersion anomaly has been detected even in the cosmic-ray protons with wave lengths of $10^{-26}m$, we expect $d^2\boldsymbol E(T_1)\ll (10^{-26} )^2m^2$ so the space appears smooth for those protons. Applying these values to $(5.8)$ and $(5.9)$, one finds the volume of ${ c}_{X_m}$ at the big bounce to be $d^3 \boldsymbol E^{3/2}(T_c)\approx 10^{-156} m^3$ compared with the Planck volume $V_p\approx 10^{-105} m^3$. Therefore, to the $O(\hbar^0)$ approximation, the big bounce does not happen at the large volume region as one might worry. On the contrary, the higher order contributions in $\hbar$ for this case are obviously important near the big bounce given by the $O(\hbar^0)$ contribution. 

 The source of these higher order contributions contains not only quantum fluctuations of the scale factor, but also inhomogeneous and anisotropic quantum fluctuations. Clearly, we may calculate these corrections only if we explicitly construct the coherent state $|\Psi\rangle$. This is an important task for the development of the model, and hopefully it will yield further insights for loop quantum cosmology.

\section{Conclusion}

This paper starts from the kinematical Hilbert space of loop quantum gravity, which describes the matter fields living in the dynamical quantum geometry of space. Using the model with a modified Hamiltonian constraint operator, we see that the dynamics of such a system reproduces FRW cosmology in the large scale limit. Further, the $O(\hbar^0) $ corrections of the model for FRW cosmology conform with loop quantum cosmology in a specific scheme. Such a result is valuable, since it attributes the predictions of loop quantum cosmology to the fundamental principles in loop quantum gravity. 

The result serves as a starting point to many possible future projects. First, one may explicitly construct the coherent states in the model to evaluate the emergent cosmology beyond $O(\hbar^0)$, to get the quantum fluctuation corrections in the emergent cosmology. Second, one may try to derive more of the implications of loop quantum cosmology by applying the model to more realistic cosmological settings. Third, one may try to improve the model by incorporating the graph-topology changing feature in the Hamiltonian constraint operator, in the hope of deriving loop quantum cosmological models with $\mu=\bar\mu$.

\section{Acknowledgments}
I would like to express gratitude to my advisor Prof. Steven Carlip, who saved no effort helping me in my research and in the preparation of this paper. This work was supported in part by Department of Energy grant DE- FG02- 91ER40674.

\end{document}